# Survey of Extra-Low Frequency and Very-Low Frequency Magnetic Fields in Cell Culture Incubators

Dong Su, Paul Héroux

*InVitroPlus Laboratory, Department of Surgery, Royal Victoria Hospital, Montreal, Canada.*

Corresponding author:
  Paul Héroux, PhD
  1110 Pine Ave West, Room 307
  Montreal, PQ, Canada
  H3A 1A3
  E-mail: paul.heroux@mcgill.ca
  http://www.invitroplus.mcgill.ca/

## Abstract

A typical cell culture $CO_2$ incubator was probed in detail to document the pattern of 60-Hz magnetic fields (MFs) inside the unit, as well as the ability of the incubator to attenuate environmental MFs. Subsequently, a survey of 46 cell culture incubators was performed. The survey measured MFs outside and inside the incubators, the frequency spectrum between 5 and 2000 Hz, and variations over time of the 60-Hz MF. Our measurements show an uneven spatial distribution, reflecting electronic and electrical components hidden within the walls. Attenuation of environmental MFs varied between 18 % and 33 %, signalling easy penetration into the units. MF levels, frequency spectra and variations over time were very different from one unit to the next. All 46 incubators surveyed had an average field greater than 0.2 µT; among them, 39 (85 %) had an average field greater than 1 µT. There is substantial work to be done in improving control over the MF environment of *in vitro* experiments in bio-medicine, particularly if they involve cancer cells.

# 1 Introduction

## 1.1 Purpose of the Survey

*In vitro* experiments are essential tools in modern toxicology, biology, and medicine. To insure meaningful *in vitro* experiments, it is necessary to tightly control the micro-environment inside incubators, for example, temperature, humidity, $CO_2$, and even $O_2$. The aim is to provide a good simulation of the *in vivo* environment, and to achieve reliable and reproducible results. However, one important factor has been overlooked for a long time by nearly all incubator manufacturers and bioscience researchers: the Magnetic Fields (MFs) within incubators, mostly generated by heating elements and control circuits.

Over the past two decades, many studies have revealed that MFs significantly affect cellular gene expression, growth and differentiation. Various cells, mostly cancer cells or chicken embryos, were used to conduct these experiments. Table 1 lists, according to ascending MFs, a number of such published findings.

Even though Table 1 is not an exhaustive list of publications on the biological effects of MFs, it shows clearly the reproducibility of some findings. For example, perturbations of the growth of MCF-7, a breast cancer cell line isolated in 1970 from a 69-year-old Caucasian woman, were confirmed by three different authors (Liburdy (1993), Harland (1997), and Blackman (2001)) in the range of 0.24 µT to 1.2 µT. Moreover, a fourth author, Girgert (2010), confirmed that MCF-7's gene expression was altered by MFs in the same intensity range. It should also be pointed out that beside this direct evidences from cell biology, several epidemiology studies have provided positive correlations between exposure to MFs and elevated risks of cancers. IARC stated that "Extremely low-frequency MFs are possibly carcinogenic to humans (Group 2B)" in its 2002 summary report. It is therefore reasonable to deduce that MFs influence biological materials.

In 2012, Li and Héroux reported that 5 cancer cells lines lost chromosomes after exposure to Extra-Low-Frequency (ELF) MFs within the range of 0.025 to 5 µT for 6 days. Further, they proposed that the underlying biological mechanism for this phenomenon is related to "an alteration in the structure of water that impedes the flux of protons in ATP Synthase channels.".

Although many researchers have pointed to significant impacts of MFs on the results of *in vitro* experiments, this critical issue has been overlooked by both incubator manufacturers and the large majority of bioscience researchers. This oversight will at the very least introduce extra variations in experimental outcomes performed in incubators with uncontrolled MFs. The situation is particularly dire in that MFs vary not only from one unit to the next, but also at different locations within the same incubator.

**Table 1.** Effects of ELF MFs (50 to 75 Hz) on experimental models.

| First Author | Year | MF (µT) | Frequency (Hz) | Study Subject | Result |
|---|---|---|---|---|---|
| Liburdy | 1993 | **0.24** | 60 | MCF-7 | Block melatonin's oncostatic action on ER+ breast cancer cell (MCF-7) proliferation |
| Harland | 1997 | **0.2-1.2** | 60 | MCF-7 | 18 % inhibition at 0.2 µT, to 15 % enhancement of growth at 1.2 µT |
| Blackman | 2001 | **1.2** | 60 | MCF-7 | Inhibition of MCF-7 cell growth |
| Girgert | 2010 | **1.2** | 50 | MCF-7 | Increased expression of p53 and c-myc |
| Berman | 1990 | **1.3 (pulse peak)** | Pulsed MF (100 pulses/s) | Chicken embryos | Increased incidence of abnormal embryos |
| Chen | 2000 | **1-1000** | 60 | Friend erythro-leukemia cells | Inhibited differentiation, 20 % at 2.0 µT, 40 % at 4.0 µT |
| Moses | 1992 | **4** | 60 | Early Chicken Embryos | Significant changes in levels of 5'nucleotidase, acetylcholinesterase and alkaline phosphatase |
| Opler | 1997 | **8** | 60 | PC12 cells | Dopamine levels significantly reduced |
| Tokalov | 2004 | **10** | 50 | Human myeloid leukemia (HL-60) cells | Significant induction of the genes HSP70A, HSP70B, and HSP70C |
| Morelli | 2005 | **73** | 75 | Membrane associated enzymes | Decrease in enzymatic activity |
| Zimmerman | 1988 | **100** | 60 | Sea Urchin | Significant developmental delay |

In 2009, Kjell Hansson Mild et al. conducted a smaller survey which investigated 6 cell incubators powered at 230 V. They reported a wide range of ELF MFs, from 0.1 µT to 38 µT, within a preset measurement grid inside the incubators. In addition, they found a maximum reading of 80 µT in the vicinity of a fan motor. Furthermore, beside concluding that "the difference in MF may very well be one of the reasons for the difference in (experimental) outcome", they pointed out that in North America, where incubators are operated at 110 V, the situation would be worse, since the current is roughly doubled.

Following these observations, it seemed important to provide a clear picture of MFs inside incubators operated in North America, and to make incubator manufacturers and users aware of this issue.

To reach these goals, we conducted a study, followed by a survey, which focused on Extra-Low Frequency and Very Low Frequency MFs, and the following items.

- Average MF in various incubators.
- Spatial distribution of MFs within incubators.
- MF variations over time inside incubators.
- Ability of incubators to attenuate environmental MFs.
- Influence of background MFs in laboratories on conditions inside the units.
- Ways to minimize the impact of MFs on experiments.

## 1.2  Types of Incubators

Generally, incubators can be assigned into 3 classes: $CO_2$ incubators, general purpose incubators (non-$CO_2$ incubators), and incubator shakers.

A $CO_2$ incubator injects $CO_2$ into its chamber, and controls $CO_2$ concentration (usually at 5%) for better control of pH within the culture medium. It is ideal for culture of mammalian cells and tissues, and has become the most popular type in medical research. $CO_2$ incubators can be further classified under "water jacket" and "air jacket" categories, according to the medium contained within their double walls. Water jacket incubators use water instead of air to heat the inner walls of incubators, and thus provide smaller temperature fluctuations compared to air jacket units. Many $CO_2$ incubators have HEPA filter systems with forced air circulation installed at the top of their chambers (Figures 1 and 2).

General purpose incubators are an economic choice for labs in other areas of study, such as microbiology and biochemistry. Unlike $CO_2$ incubators, most general purpose incubators are air jacketed, and without forced air circulation, to reduce cost. Without $CO_2$ injection, circulation, and concentration control components, and without the powerful motors required to circulate air through HEPA filters, we expected that that they would have weaker MFs than $CO_2$ incubators. Some studies on mammalian cells can be performed in general purpose incubators by substituting a soluble buffer (often HEPES) in the medium for $CO_2$.

An incubator shaker is an incubator with a mechanical shaker platform (Figure 3). The main application is for the growth of bacterial culture.

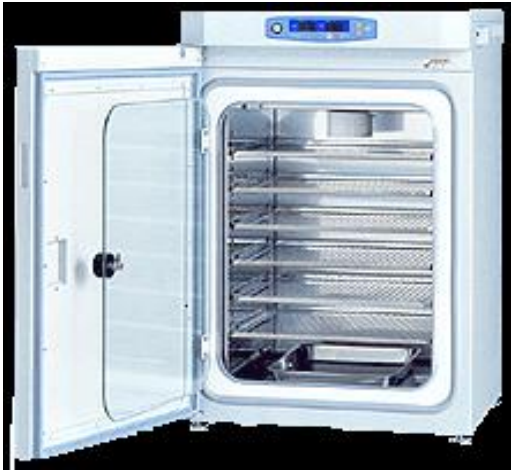

***Figure 1.*** CO$_2$ *incubator.*

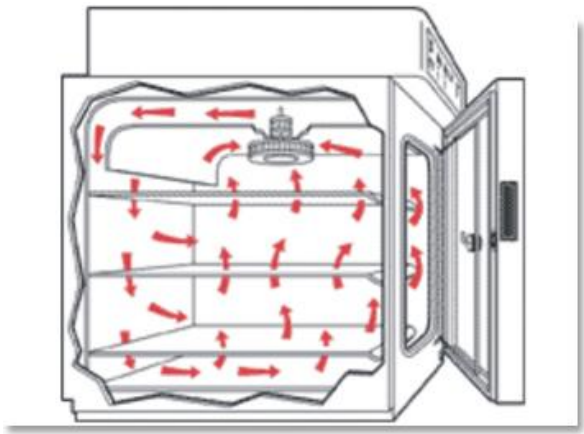

***Figure 2****. HEPA Filter system with forced air circulation.*

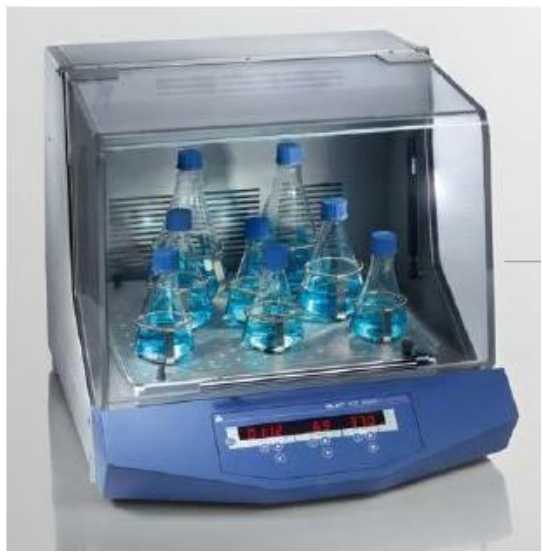

***Figure 3****. Shaker incubator.*

# 2 Materials and Methods

## 2.1 MF Meters

Two MF measuring instruments were used in this study. The first was a Field Star 1000 (Dexsil Corporation, Hamden, CT, USA), capable of recording MFs in X-Y-Z axes, and their vectorially integrated RMS value each second, within a narrow bandwidth of 55 to 65 dB ±3 dB. This meter was exclusively used for recording the time variations of 60-Hz MFs. A second survey meter, a Narda EFA-300 (Narda Safety Test Solutions, Pfullingen, Germany), comes with a detachable probe connected with a cable. The Narda has the ability to measure the spectrum of MFs from 5 Hz to 32 kHz, and field strengths ranging from 100 nT to 32 mT.

The exact bandwidth on the Narda is selectable (for example, 5 Hz to 2 kHz), but the measurements of the two instruments are compatible, because energy is tightly concentrated around 60-Hz. Both meters were calibrated using a 1 meter square coil and a Fluke 5100B calibrator.

The Field Star has a narrow bandwidth, but is capable of recording 60-Hz MF variations more quickly. Therefore, we used it during the survey to record changes in 60-Hz MFs over time. The Narda EFA-300 has a wider bandwidth, but the recording interval is not less than 5 seconds. Therefore, we chose the Field Star to record time variations, while using the Narda EFA-300 to record MFs over a 5 Hz to 2 kHz bandwidth, as well as the frequency spectrums.

## 2.2 Detailed study of internal MFs

In order to get a full picture of the distribution of MFs inside incubators, we selected two incubators (Baxter WJ501 and Thermo Forma 3310) for detailed study. We setup sampling grids, and measured at each grid point. In order to trace the origin of the various MF contributions, we removed all the covering panels of the Baxter WJ501, except for the bottom one, exposing many electrical components. Measurements were performed twice, with separate measurements for heater "on" and "off".

Incubators usually have a glass door, which allows viewing of the inside, covered by a second opaque door which is normally kept closed, except during observations. This second door is heated to avoid condensation of humidity on the glass door, so that transparency of the glass is maintained. The MF contribution of the heating wire located inside the second door was separately assessed using the same measurement grid.

## 2.3  MF Attenuation by Incubator Walls

To measure the Baxter WJ501's (exterior dimensions of 74 by 64 by 74 cm) ability to attenuate environmental MFs, we constructed a 238 cm X 238 cm square coil, and used it to apply a controlled external MF around the unit. The coil was successively placed in 3 different orthogonal directions, to allow measurement of the attenuation of MFs oriented from top to bottom (Z), door to back (Y), and side to side (X). Because current cannot circulate across the door opening, even when the door is closed, it is not expected that attenuations will be the same in all field orientations. These attenuation measurements are difficult to perform, due to the space necessary to setup the coils, and to the difficulty and safety concerns involved in positioning a heavy water jacketed incubator within the magnetic coils. Even a relatively small incubator weighs about 200 kg, when water-loaded. Although we provide in this document measurements of attenuation for one incubator only, we feel that this single measurement is representative, based on the fact that most incubators have similar stainless steel double-wall construction. This type of measurement has never been performed before. The Narda EFA-300 was used for these measurements.

### 2.3.1  Z-axis

For the Z (vertical) axis, we centered the horizontal coil vertically on the incubator (Figure 4b), and applied a fixed current (60-Hz) to create a MF. We measured at 9 points (A to I, Figure 4a) in a horizontal plane at mid-height inside the incubator, with the incubator door tightly closed, using the probe of the Narda EFA-300.

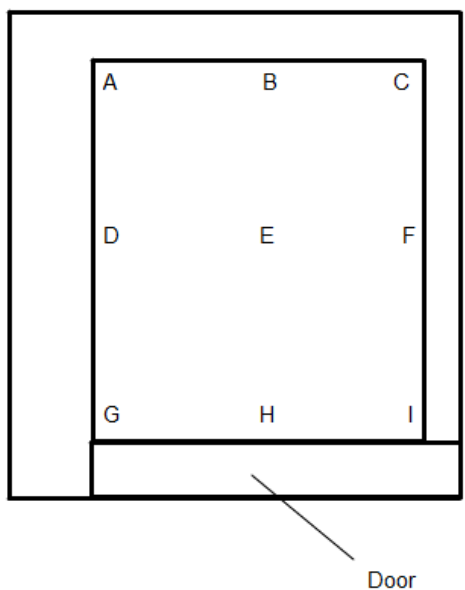

***Figure 4a***.    *2D plan view of the 9 measurement points (A-I) in the Baxter WJ501 incubator.*

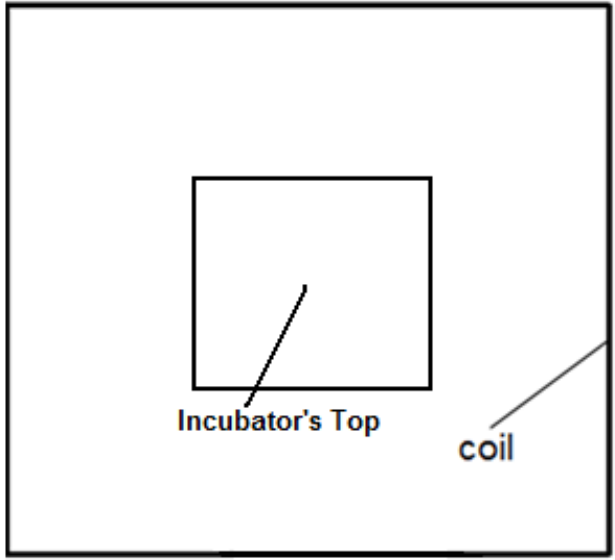

***Figure 4b.*** *2D plan view of the Baxter WJ501 incubator, placed at the center of the coil, for the measurement of Z-axis attenuation.*

Although the cable of the probe has a diameter of about 8 mm, it is still easy to close both layers of doors tightly over the cable, since cell incubators are fitted with rubber door seals, usually thicker than 10 mm. Thus, the inside glass door can still be closed and tightened with the latch. In addition, the outer door is also equipped with an even thicker rubber seal, which allows passage of the probe's cable without affecting the closure of the door.

In order to measure the door's contribution to MF attenuation, we then fully opened the door, and measured at the same points again. For the purpose of getting reproducible readings, we set the meter to RMS and "Max" modes. After these measurements, we removed the incubator from the MF, and measured the fields at the same locations, to obtain the MFs without the attenuation due to the incubator. All functions of the incubator were "off" during these measurements.

### 2.3.2 Y-axis

For the Y-axis (door to back), we placed the coil vertically, parallel to the back of the incubator (Figure 5), and measured at the same 9 points (Figure 4a).

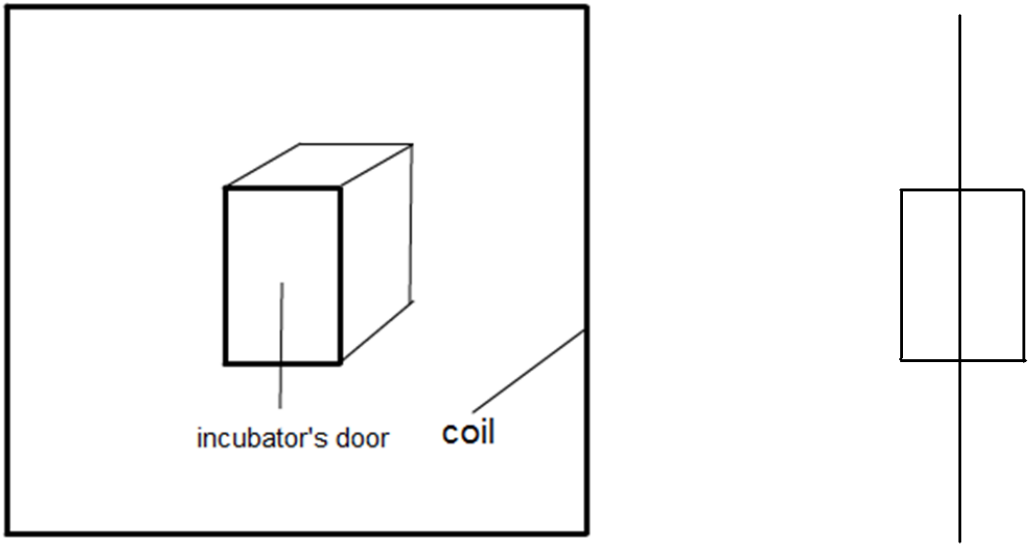

***Figure 5.*** *Square MF coil placed parallel to the back of the incubator.*
*Left: 3D front view.   Right: 2D side view.*

### 2.3.3   X-axis

For the X-axis (side to side), we placed the coil vertically, parallel to the sides of the incubator (Figure 6), and measured the same points (Figure 4a).

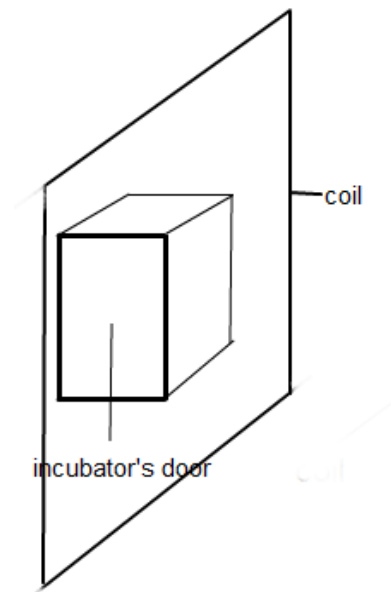

***Figure 6.***  *3D front view of the square MF coil placed parallel to the side walls of the incubator.*

## 2.4  MF Survey Procedures

Having completed these measurements (2.2 and 2.3), we performed our main survey, conducted in 14 different departments of the Royal Victoria Hospital and McGill University laboratories (Faculties of Medicine and Biology, Cancer Research Center). 87 incubators were indentified, and 46 of them were measured. Since the units surveyed were actively in use, the following procedures take into account encumbrances due to the presence of other pieces of laboratory equipment around the incubators, and of positioning of culture vessels within incubators. The measurements were performed with the Narda EFA-300, except for 2.4.2e.

### 2.4.1.  MF Measurements Outside the Incubators

a) Measure the MF in front of the centre of the incubator's door, at a distance of 1 meter; measurement time is 1 minute, using RMS Max mode at 60-Hz.

b) Measure the 60-Hz MF at the left, right, and back, at a distance of 1 meter of the respective centers on the external surfaces of the incubator. If there is any other electrical device within 1 meter, measure at 50 cm. When the space between two units is less than 50 cm, measure at midway between the incubator and the device.

### 2.4.2  MF Measurements Inside the Incubators

a) All the measurements use Max mode, measurement times are at least 1 minute after the reading stabilizes, in order to catch the peak value (when the heater is on). If there are fewer than 3 layers of shelves installed in the incubator, measure at height/4, height/2, and 3x height/4. For each height, measure at 5 cm from the left wall, at the mid-point, and at 5 cm from the right wall.

c) If there are 3 or more layers of shelves inside the incubator, measure at the shelves closest to the height/4, height/2, and 3x height/4. For each shelf, measure at 5 cm from the left wall, middle point, and 5 cm from the right wall.

d) At the center point of the incubator, record the MF's frequency spectrum, using the Narda EFA-300's Vect Max mode.

e) At the center point of the incubator, using the Field Star 1000, record the MF (60-Hz) at each second for a duration of 5 minutes, or the time for 10 cycles of heater "on" and "off", whichever is longer.

f) For incubator shakers, measure the 3 columns of the tray: left, middle, and right columns. At each column, measure at 3 different points: front, middle and rear holders. Frequency spectrum and time variations of MFs are also recorded, as in steps d) and e).

# 3   Results

We define 3 incubator axes to report our MF vector measurements: "X-axis" from left to right, "Y-axis" from front to back, and "Z-axis" from bottom to top. For each axis, we plotted the Narda readings from grids at a series of fixed heights. The vectorially integrated MFs (RMS of combined X-Y-Z) are also reported.

The Field Star 1000 logged X-Y-Z MF components, as well as the integrated values, while recording MFs over time (section 3.2.2.6).

## 3.1 Detailed MF Measurements

One of the incubators available for detailed measurements (Baxter) could be disassembled to localize the electrically powered components, while the second one (Forma) was a larger unit that we used for comparison.

### 3.1.1   MFs inside the Baxter Incubator

We made the following measurements in the Baxter WJ501 incubator,

#### 3.1.1.1   MFs along X-axis

MFs along the X-axis (left to right) increase along height (Z axis), as can be seen from the summary statistics in Table 2, and the plots in Appendix A.

**Table 2.** Mean, minimum, maximum, and standard deviation of MFs along the X-axis at different heights inside the Baxter WJ501.

| Height from Bottom (Z, cm) | MFs along the X-axis (µT) | | | |
|---|---|---|---|---|
| | **Mean** | **Min** | **Max** | **Std Dev** |
| 0 | 0.61 | 0.07 | 1.00 | 0.22 |
| 7.3 | 0.63 | 0.25 | 1.20 | 0.25 |
| 14.6 | 0.67 | 0.04 | 1.24 | 0.26 |
| 21.9 | 0.77 | 0.21 | 1.56 | 0.39 |
| 29.2 | 0.99 | 0.14 | 2.64 | 0.80 |
| 36.5 | 1.84 | 0.24 | 6.20 | 1.9 |

#### 3.1.1.2   MFs along Y-axis

MFs along the Y-axis (front to back), from Table 3 and the plots in Appendix B, showed the following.

First, at the bottom of the incubator (0 cm height), the intensity of the MFs along the Y-axis display a very clear wavy pattern (Figure 7), which could be explained by a heating element arrayed along the X-axis, underneath the bottom panel. This makes design sense in terms of spreading out the heating power.

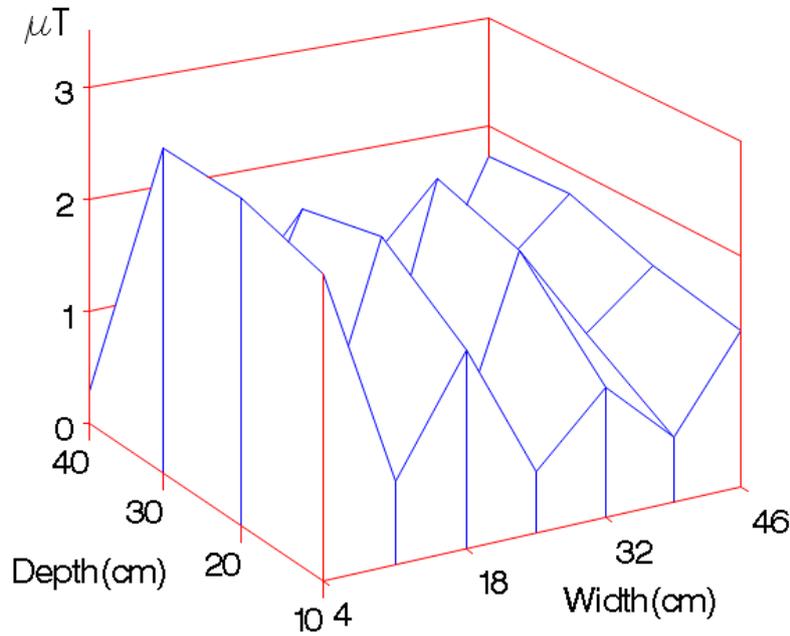

*Figure 7.*   *Wavy MF pattern at the bottom of Baxter WJ501.*

Second, we noted that as we moved towards the top of the incubator, the average MF along the Y-axis initially decreased, and then increased (Table 3). This was mainly caused by an air circulation motor, which is located at the left rear top of the incubator.

**Table 3.** Mean, minimum, maximum, and standard deviation of MFs along the Y-axis at different heights inside the Baxter WJ501.

| Height from Bottom (Z, cm) | MFs along the Y-axis (µT) | | | |
|---|---|---|---|---|
| | Mean | Min | Max | Std Dev |
| 0 | 1.24 | 0.05 | 2.84 | 0.82 |
| 7.3 | 0.51 | 0.06 | 1.76 | 0.49 |
| 14.6 | 0.48 | 0.04 | 2.58 | 0.61 |
| 21.9 | 0.30 | 0.06 | 1.52 | 0.35 |
| 29.2 | 0.36 | 0.03 | 1.44 | 0.29 |
| 36.5 | 0.77 | 0.13 | 2.24 | 0.46 |

In addition, we noted that at a height of 14.6 cm, in the left front corner, the MF peaked (Figure 8), caused by a circuit board with a transformer (Figure 20) within the wall.

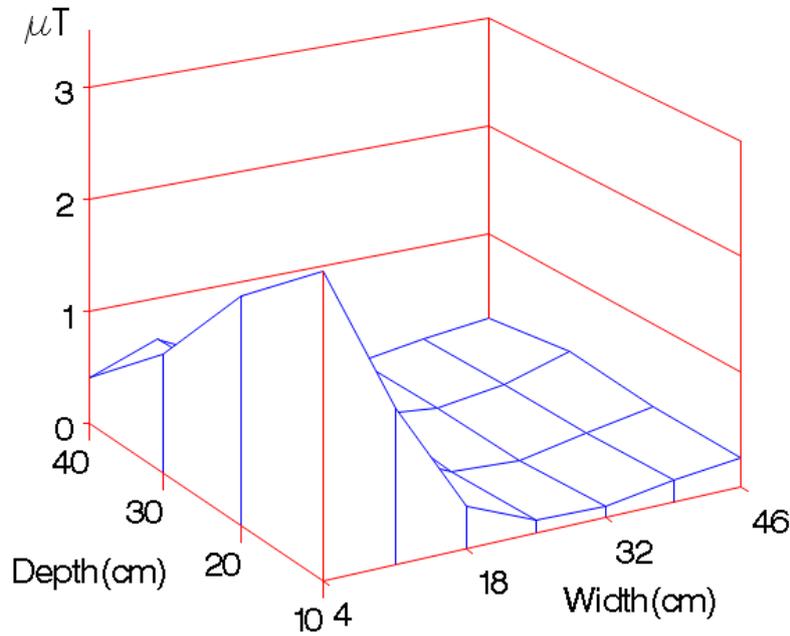

*Figure 8.*   MF along the X-axis peaked near a Transformer.

### 3.1.1.3   MFs along Z-axis

MFs along the Z-axis (bottom to top) increase substantially with height, from summary Table 4, and the plots in Appendix C.

**Table 4.** Mean, minimum, maximum, and standard deviation of MFs at different heights inside the Baxter WJ501 along the Z-axis.

| Height from Bottom (Z, cm) | MFs along the Z-axis (µT) | | | |
|---|---|---|---|---|
| | **Mean** | **Min** | **Max** | **Std Dev** |
| 0 | 0.93 | 0.25 | 1.84 | 0.43 |
| 7.3 | 0.34 | 0.16 | 0.62 | 0.13 |
| 14.6 | 0.31 | 0.08 | 0.78 | 0.18 |
| 21.9 | 0.72 | 0.06 | 0.36 | 0.36 |
| 29.2 | 1.16 | 0.11 | 2.48 | 0.78 |
| 36.5 | 3.48 | 0.06 | 11.2 | 3.49 |

The plots in Appendix C indicate a rise up to 11.2 µT along the Z-axis (Figure 9), which was caused by the air circulation motor previously mentioned. The Z component was dominant.

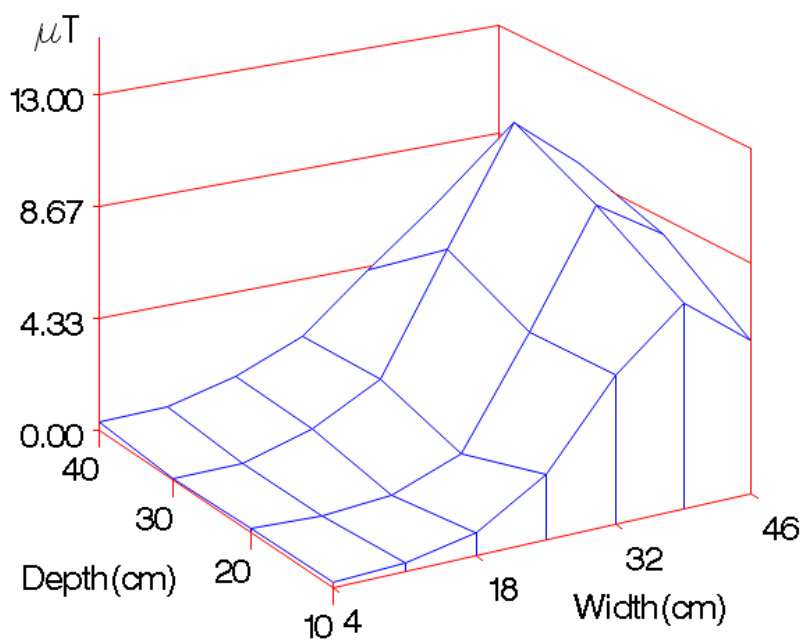

***Figure 9***.   *MF along the Z-axis peaks under a fan motor.*

### 3.1.1.4   Integrated MFs

Finally, the results of the integrated (X-Y-Z together) MF (Table 5 and Appendix D). The Z-component dominates the integrated MF results.

**Table 5.** Integrated mean, minimum, maximum, and standard deviation of MFs at different heights inside the Baxter WJ501.

| Height from Bottom (Z, cm) | Integrated MFs (µT) | | | |
|---|---|---|---|---|
| | Mean | Min | Max | Std Dev |
| 0 | 1.86 | 0.8 | 3.40 | 0.74 |
| 7.3 | 0.95 | 0.40 | 2.36 | 0.45 |
| 14.6 | 0.99 | 0.31 | 2.84 | 0.53 |
| 21.9 | 1.22 | 0.53 | 2.44 | 0.44 |
| 29.2 | 1.74 | 0.52 | 3.44 | 0.90 |
| 36.5 | 4.32 | 0.42 | 12.80 | 3.74 |

### 3.1.1.5   MF contribution of the Door

Inside their exterior door, most cell incubators contain a heating element to warm the inside glass door. The power dissipated by this heating element is often adjustable. In order to evaluate the maximum influence of the door's MF, we adjusted the door heater to the 'Max' setting, and conducted MF measurements inside the incubator. The results are shown in Table 6 with detailed plots in Appendix E.

**Table 6.** Mean, minimum, maximum, and standard deviation of MFs with maximum door heating at different heights inside the Baxter WJ501.

| Height from Bottom (Z, cm) | Integrated MFs (µT) | | | |
|---|---|---|---|---|
| | Mean | Min | Max | Std Dev |
| 0 | 0.09 | 0.04 | 0.20 | 0.04 |
| 7.3 | 0.09 | 0.04 | 0.22 | 0.05 |
| 14.6 | 0.10 | 0.05 | 0.25 | 0.05 |
| 21.9 | 0.12 | 0.06 | 0.29 | 0.06 |
| 29.2 | 0.13 | 0.06 | 0.41 | 0.07 |
| 36.5 | 0.15 | 0.06 | 0.56 | 0.10 |

The peak readings (Figure 10) at the front left corner (0.56 µT) resulted from the wiring connection to the door heater. It was easy to expose this wiring, and a better design for the heating wire itself would return current by a route folded on itself, to reduce the MFs (the actual wiring is shown in Figure 19).

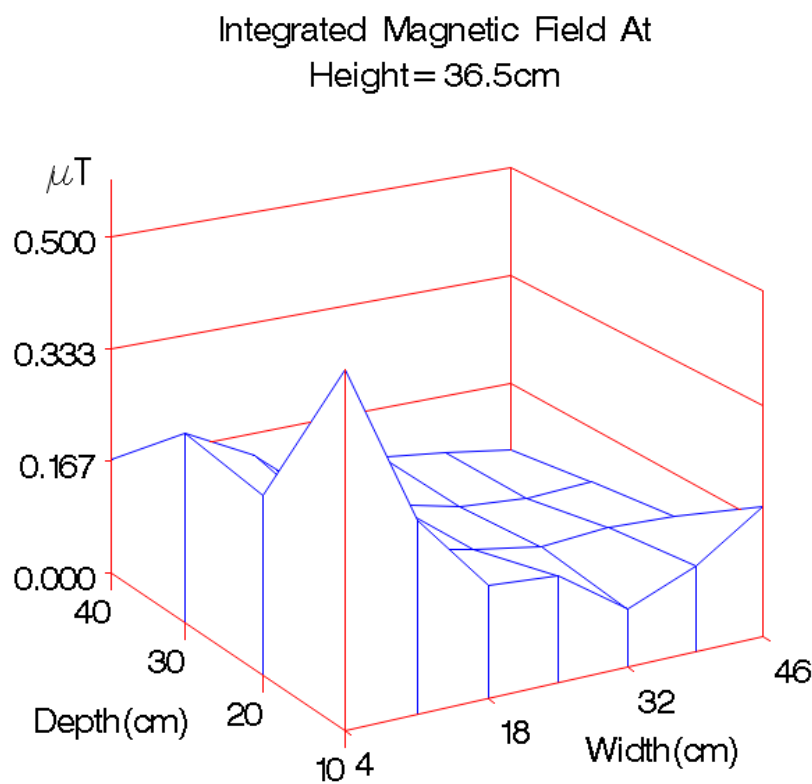

*Figure 10.  Wire connection between incubator door and body produces a peak MF at the left front corner.*

By comparing MFs at the same locations with the door opened and closed, we can ascertain the contribution of the door to the incubator's total field. From Table 7, we can conclude that the contribution of the door is relatively minor, somewhere between 3.5 % and 10.7 % of the total MF, for this particular incubator.

**Table 7.** Door Contribution to the MF at different heights inside the Baxter WJ501.

| Height (Z, cm) | Mean with Door open (µT) | Mean with Door closed (µT) | Door Contribution |
|---|---|---|---|
| 0 | 0.094 | 1.86 | 5.05% |
| 7.3 | 0.092 | 0.954 | 10.7% |
| 14.6 | 0.104 | 0.986 | 10.5% |
| 21.9 | 0.120 | 1.224 | 9.80% |
| 29.2 | 0.133 | 1.737 | 7.66% |
| 36.5 | 0.153 | 4.323 | 3.5% |

### 3.1.1.6 MFs with Heating "Off"

Heater "off" measurements isolate the contribution of all non-heating incubator systems. We list the integrated MFs only in Table 8, with complete plots in Appendix F.

**Table 8.** MFs inside the Baxter WJ501, when heater is "off".

| Height from Bottom (Z, cm) | Integrated MFs (µT) | | | |
|---|---|---|---|---|
| | Mean | Min | Max | Std Dev |
| 0 | 0.51 | 0.32 | 1.04 | 0.17 |
| 7.3 | 0.68 | 0.41 | 1.48 | 0.29 |
| 14.6 | 0.89 | 0.44 | 2.36 | 0.42 |
| 21.9 | 1.08 | 0.45 | 1.80 | 0.36 |
| 29.2 | 1.71 | 0.51 | 3.32 | 0.88 |
| 36.5 | 3.54 | 0.53 | 8.88 | 2.60 |

With the heater off, the wavy pattern (due to the heating element) at the bottom of the incubator disappears. Ratios of integrated MFs between heater "on" and "off" are shown in Table 9.

**Table 9.** MF ratios between heater "on" and heater "off" in the Baxter WJ501.

| Height (Z, cm) | Mean with Heater "off" (µT) | Mean with Heater "on" (µT) | Mean "on" / Mean "off" |
|---|---|---|---|
| 36.5 | 3.541 | 4.323 | 1.22 |
| 29.2 | 1.705 | 1.737 | 1.02 |
| 21.9 | 1.085 | 1.224 | 1.13 |
| 14.6 | 0.887 | 0.986 | 1.11 |
| 7.3 | 0.679 | 0.954 | 1.41 |
| 0 | 0.51 | 1.86 | 3.65 |

From Table 9, the heater contributed most of the MF in the lower part of the incubator, while it did not make a significant contribution in the upper part of the incubator, where the air circulation motor was dominant.

### 3.1.1.7 Attenuation of Environmental MFs

The double-walled structure of welded stainless steel generally used in incubators should provide some attenuation of external MFs, because of its electrical conductivity, even though stainless steel is commonly non-magnetic. As these attenuations are expected to be different for each axis, tables are presented for X, Y and Z orientations.

#### 3.1.1.7.1  X-axis (left to right)

**Table 10 .** External X-axis MF attenuations by Baxter WJ501.

| Location (Figure 4) | MF without incubator (µT) | MF with incubator door closed (µT) | MF with incubator door open (µT) | Attenuation with door closed (%) |
|---|---|---|---|---|
| A | 9.648 | 7.144 | 7.267 | 25.95 |
| B | 11.3 | 8.327 | 8.693 | 26.31 |
| C | 12.86 | 9.293 | 9.64 | 27.74 |
| D | 8.831 | 6.613 | 6.818 | 25.11 |
| E | 10.86 | 7.992 | 8.303 | 26.41 |
| F | 12.74 | 9.444 | 9.931 | 25.87 |
| G | 9.32 | 8.437 | 9.252 | 9.47 |
| H | 11.28 | 8.626 | 9.443 | 23.52 |
| I | 12.69 | 10.49 | 13.13 | 17.33 |
| **Average** | **11.05** | **8.49** | **9.16** | **23.08** |

### 3.1.1.7.2 Y-axis (front to back)

Table 11. External Y-axis MF attenuations by Baxter WJ501.

| Location (Figure 4) | MF without incubator (µT) | MF with incubator door closed (µT) | MF with incubator door open (µT) | Attenuation with door closed (%) |
|---|---|---|---|---|
| A | 11.7 | 7.876 | 7.909 | 32.68 |
| B | 11.91 | 8.513 | 8.885 | 28.52 |
| C | 12.13 | 8.847 | 9.211 | 27.06 |
| D | 11.19 | 6.932 | 7.388 | 38.05 |
| E | 11.12 | 7.087 | 8.238 | 36.27 |
| F | 11.02 | 8.072 | 8.682 | 26.75 |
| G | 9.87 | 6.675 | 7.580 | 32.37 |
| H | 9.5 | 5.147 | 8.23 | 45.82 |
| I | 9.99 | 7.812 | 8.505 | 21.08 |
| **Average** | **10.94** | **7.44** | **8.29** | **32.15** |

### 3.1.1.7.3 Z-axis (bottom to top)

Table 12. External Z-axis MF attenuations by Baxter WJ501.

| Location (Figure 4) | MF without incubator (µT) | MF with incubator door closed (µT) | MF with incubator door open (µT) | Attenuation with door closed (%) |
|---|---|---|---|---|
| A | 13.06 | 10.06 | 10.09 | 22.97 |
| B | 12.62 | 10.4 | 10.46 | 17.59 |
| C | 13.13 | 11.44 | 11.48 | 12.87 |
| D | 13.34 | 9.82 | 9.86 | 26.39 |
| E | 12.24 | 9.77 | 9.9 | 20.18 |
| F | 13.10 | 10.97 | 11.12 | 16.25 |
| G | 13.16 | 9.87 | 10.53 | 25 |
| H | 12.25 | 10.25 | 10.81 | 16.32 |
| I | 13.03 | 11.45 | 11.65 | 12.13 |
| **Average** | **12.88** | **10.44** | **10.65** | **18.86** |

### 3.1.1.7.4 Conclusion

From Tables 12, 13 and 14 above, environmental MFs are attenuated most when applied from back to door of the incubator, followed by fields applied from left to right, and finally from top to bottom. The average attenuations vary between 18 % and 33 %.

### 3.1.2 MFs of the Forma 3310 Incubator

It is not expected that the MF patterns inside incubators of different makes and models will be closely matched, in spite of generally similar structures, because of different shapes and volumes, and because of the different placement of active components. For the two incubators which were investigated in detail, we compared heating "on" MFs at various levels. The most evident difference was that the Forma incubator had much weaker fields than the Baxter incubator, this being partly due to the larger internal volume of the Forma (150 vs 315 liters). The recorded ratios at corresponding heights are shown in Table 11, with plots in Appendix G.

**Table 13.** Distribution of MFs inside the Forma 3310, with heater "on".

| Height from Bottom (Z, cm) | Integrated MFs (µT) | | | |
|---|---|---|---|---|
| | Mean | Min | Max | Std Dev |
| 5 | 0.23 | 0.1 | 0.56 | 0.13 |
| 15 | 0.31 | 0.11 | 0.84 | 0.19 |
| 25 | 0.43 | 0.14 | 1.12 | 0.29 |
| 35 | 0.57 | 0.16 | 1.56 | 0.42 |
| 45 | 0.73 | 0.18 | 2.56 | 0.63 |
| 55 | 0.66 | 0.11 | 2.28 | 0.62 |

**Table 14.** Comparison of mean MFs between Forma 3310 and Baxter WJ501.

| Level | Mean of Forma (µT) | Mean of Baxter (µT) | Ratio Forma/Baxter |
|---|---|---|---|
| 0 | 2.28 | 18.6 | 12.3% |
| 1 | 3.1 | 9.54 | 32.5% |
| 2 | 4.25 | 9.86 | 43.1% |
| 3 | 5.71 | 12.24 | 46.7% |
| 4 | 7.29 | 17.37 | 42.0% |
| 5 | 6.65 | 43.23 | 15.4% |

The Forma incubator was more uniform at each height. At the bottom, it didn't show a wavy pattern; at the top level, it did not peak at the rear right (because of an air circulation motor). These numbers show that the internal electrical components, and their layout in the Forma, were different from the Baxter.

## 3.2 Survey of 46 Incubators

### 3.2.1 MFs of Incubator Environments

Among the 46 incubators surveyed, 35 were found to have other electrical device(s) within 1 meter. Other incubator units, biological safety cabinets, refrigerators or freezers were most often seen. The fields produced by these devices ranged from 0.07 µT to 18.9 µT, measured at their surface. Without the influence of such other electrical devices within 1 meter around the incubators, the background fields were found to be in the range of 0.033 µT to 1.28 µT. From Table 15, we can see that 41 out of 46 incubators have inside maximum MFs greater than their background fields, with a median ratio of 8.71. This supports our conclusion that MFs produced by the incubators themselves are generally dominant.

### 3.2.2 MFs inside Incubators

For each incubator, 9 readings were recorded at the different locations described in Figure 4a. The average values of the 9 readings are shown in the following table as 'Mean'. The minimum and maximum of the 9 reading are designated as 'Min' and 'Max' (Table 15).

**Table 15.** Summary MF Table of 46 Surveyed Incubators (in µT).

| Brand | Model | Type | Mean | Min | Max | Max Background |
|---|---|---|---|---|---|---|
| New Brunswick | G-25 | Shaker | **0.39** | 0.2 | 0.81 | 2.06* |
| Chicago Surgical Ele. | N.A. | General | **0.61** | 0.25 | 1.21 | 3.32* |
| Forma Scientific | 3956 | General | **0.76** | 0.2 | 2.64 | 0.22 |
| Fisher Sci. | Isotemp | General | **0.76** | 0.05 | 1.85 | 0.32 |
| Fisher Sci. | 637D | General | **0.84** | 0.22 | 2.49 | 0.23 |
| Forma Scientific | 3157 | $CO_2$ W | **0.91** | 0.11 | 2.66 | 1.77* |
| Thermo Electron | N.A. | Shaker | **0.98** | 0.57 | 1.58 | 5.86* |
| Nuaire | US auto flow | $CO_2$ W | **0.99** | 0.4 | 2.28 | 1.34* |
| Thermo Forma | 3310 | $CO_2$ W | **1.04** | 0.32 | 3.75 | 0.68* |
| Innova New Brunswick | 4200 | Shaker | **1.17** | 0.31 | 2.97 | 0.4 |
| Fisher Isotemp | 281 | General | **1.86** | 1.2 | 2.22 | 0.47 |
| Baxter | WJ501 | $CO_2$ W | **1.87** | 0.77 | 5.27 | 1.6* |
| Sanyo | N.A. | $CO_2$ | **2.77** | 0.85 | 6.72 | 0.3 |
| New Brunswick | G-25 | Shaker | **2.79** | 0.42 | 16.13 | 0.31 |
| Sanyo $O_2$/ $CO_2$ | MCO-18M | $CO_2$ | **2.8** | 1.48 | 4.14 | 0.81* |
| Sanyo | MCO_19AIC | $CO_2$ | **2.94** | 1.63 | 5.17 | 3.31* |
| Sanyo | MCO-20AIC | $CO_2$ | **3.12** | 1.22 | 6.64 | 6.68* |
| Hera Cell | 240 | $CO_2$ | **3.28** | 2.36 | 4.62 | 1.48* |
| Baxter | Tempcon | General | **3.36** | 0.61 | 7.43 | 1* |
| Innova New Brunswick | 4000 | Shaker | **3.47** | 1.27 | 9.53 | 0.36 |
| Hera Cell | N.A. | $CO_2$ | **3.65** | 2.68 | 4.49 | 0.26* |
| Thermo Scientific | 370 | $CO_2$ | **3.84** | 1.9 | 7.01 | 0.64* |
| New Brunswick | C25 | Shaker | **3.88** | 0.33 | 17.74 | 0.96* |
| Thermo Electron | 3110 | $CO_2$ W | **3.91** | 1.19 | 8.56 | 0.92* |
| Nuaire | Nu4750 | $CO_2$ W | **3.95** | 0.77 | 10.38 | 0.64* |
| Thermo Scientific | 370 | $CO_2$ | **3.99** | 2.03 | 6.25 | 0.96* |
| Forma Scientific | 3130 | $CO_2$ W | **4.67** | 1.53 | 11.14 | 1.37* |
| Forma Scientific | 3110 | $CO_2$ W | **5.44** | 1.77 | 12.59 | 2.42* |
| Fisher Sci. | 546 | $CO_2$ W | **6.58** | 2.36 | 16.88 | 0.38 |
| Forma Scientific | N.A.(Old) | $CO_2$ | **6.71** | 2.32 | 16.83 | 1.36* |
| Thermo Electron | 3130 | $CO_2$ W | **6.79** | 1.73 | 16.97 | 18.9*** |
| Thermo Electron | 3110 | $CO_2$ | **7.55** | 1.83 | 18.28 | 3.92* |
| Revco | N.A.(Old) | $CO_2$ | **7.67** | 3.57 | 17.76 | 1.27* |
| Napco | 3550 | $CO_2$ | **7.8** | 3.52 | 13.42 | 2.84* |
| Thermo Electron | Napco 3550 | CO2 | **7.83** | 3.81 | 12.13 | 1.63* |
| Fisher Sci. | Isotemp 546 | $CO_2$ W | **9.61** | 2.34 | 37.58 | 0.76* |
| Thermo Forma | 3110 | $CO_2$ W | **9.73** | 2.73 | 24.14 | 0.47* |
| N.A. | N.A. | General | **10.46** | 3.57 | 19.51 | 0.2 |
| Thermo Forma | 3110 | $CO_2$ W | **11.89** | 3.3 | 30.41 | 0.49* |
| Gallenkamp | N.A. | General | **11.96** | 3.06 | 37.17 | 2.3* |
| Fisher Sci. | 610 | $CO_2$ | **12.3** | 5.15 | 35.52 | 1.59* |
| Forma Scientific | 3158 | $CO_2$ W | **13.08** | 2.62 | 50.64 | 1.61* |
| Labline | 3527 | Shaker | **14.04** | 3.62 | 42.74 | 11.87** |
| WWR international | 2005 | General | **15.48** | 4.92 | 47.37 | 1.28 |
| Forma Scientific | 546 | $CO_2$ | **16.5** | 2.61 | 74.47 | 3.45* |
| Sanyo | MIR152 | $CO_2$ | **26.98** | 5.67 | 120 | 0.34* |

Note: Type "$CO_2$ W" means $CO_2$ incubator with water jacket.

* measured at 50 cm or halfway between the incubator and other electric equipment.

** 5 cm to another incubator.

*** 10 cm to a power outlet panel.

The above table displays a large variety of MFs, which resulted from the following factors:

1. Different makes or models are designed differently, the electrical components are different, and their layouts are also different.
2. Even identical models can have a range of fields, as a result of
    a) different periods (30-60minutes) of heating cycles, as a result of ambient conditions,
    b) parts variations within a model, specifically different electrical components,
    c) changes in design over time, under the same model number,
    d) repairs or maintenances altering electrical parts, and
    e) different background MFs.

We measured the size of each incubator, and also tried to collect information about the year of manufacture. However, we found this information could not be easily obtained, due to incomplete documentation, and more importantly, the limited enthusiasm and time of the busy incubator users. It is not feasible to move heavy incubators, while they are in use, to access the chassis plates. It was not possible to produce a regression between MFs and date of manufacture.

### 3.2.2.1  Average and Maximum MFs

From the histogram in Figure 11, we can see that the 'Mean' MFs of the 46 incubators produced a right-skewed distribution, with a median around 4 µT.

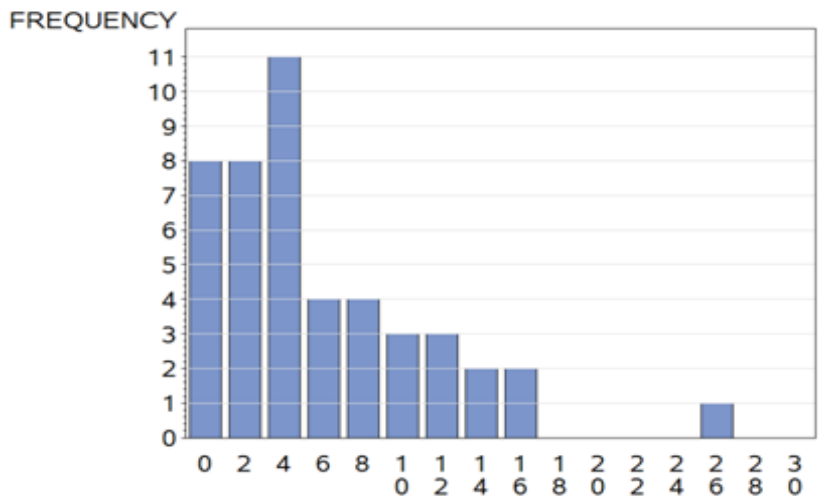

***Figure 11 .***   *Distribution of Average MFs among 46 surveyed Incubators ( µT).*

For the maximum reading ('Max', i.e. maximum field), the histogram in Figure 12 shows a pattern similar to Figure 11, but with the median of 10 µT.

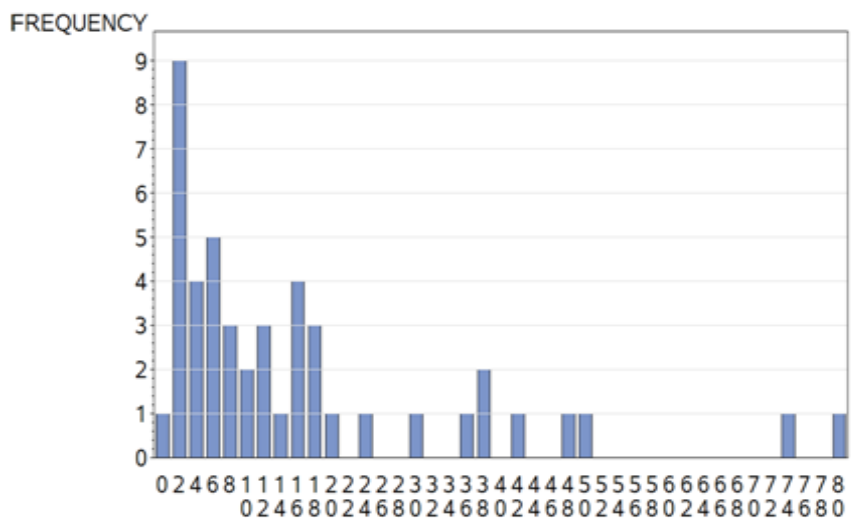

***Figure 12***.   *Distribution of Maximum MFs among 46 surveyed incubators ( µT).*

### 3.2.2.2   Average and Maximum MFs by Incubator Type

From Figures 13, we see that air jacket incubators have higher MFs than those with water jackets. This may be because water jacket incubators are designed with lower heater powers.

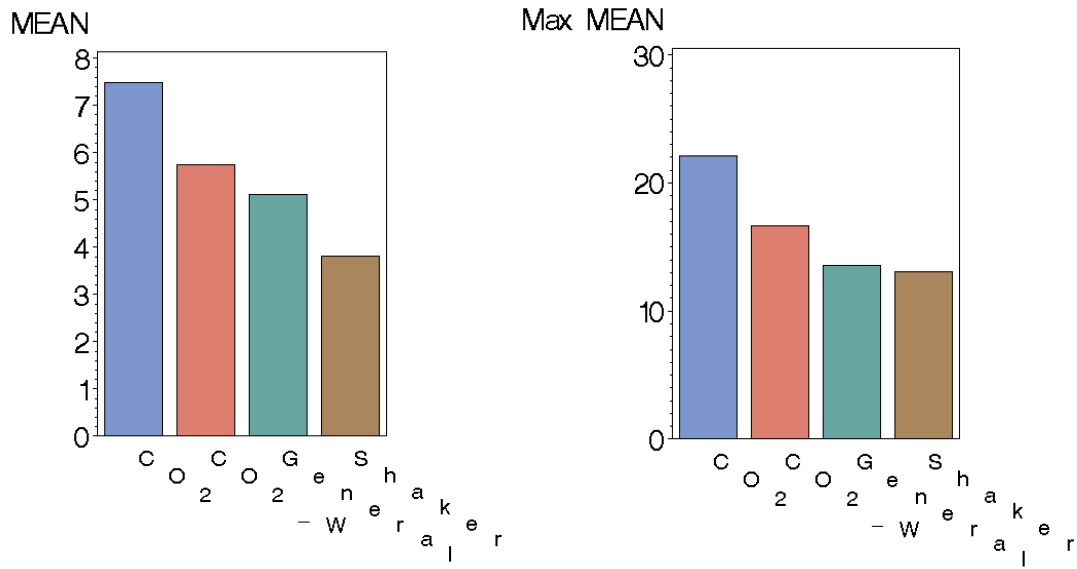

***Figure 13.***   *Left: Average MF (Mean in µT) by incubator type.*
              *Right: Average Maximum MF (Max Mean in µT) by incubator type.*

### 3.2.2.3   Location of Minimum MF

Figure 14 classifies the location of mimimum MFs for different types of incubators. The locations of minimum and maximum fields provide a generic desciption of how MFs are distributed inside most incubators, but each particular incubator must be surveyed for precise locations.

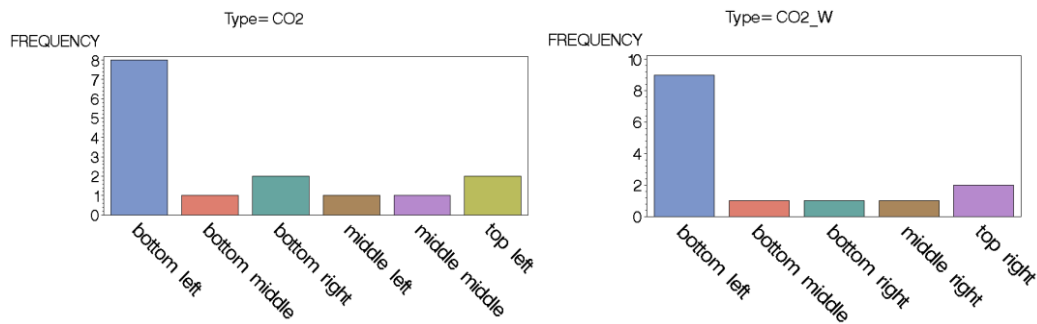

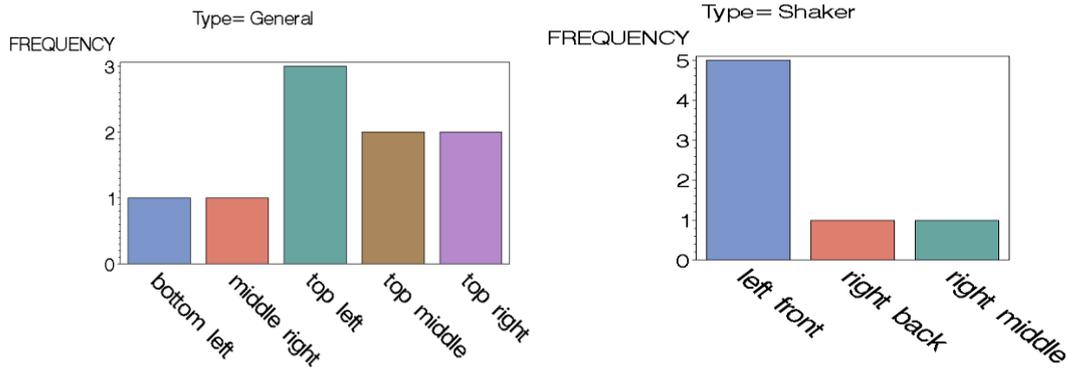

***Figure 14.*** *Location of minimum MFs, by incubator type.*
*Top left:* CO$_2$ *incubators.*
*Top right:* CO$_2$ *incubators with water jacket.*
*Bottom left: General purpose incubator.*
*Bottom right: Incubator shakers.*

Most CO$_2$ incubators, no matter whether water or air jacketed, predominately have their minimum fields at the bottom shelves, to the left hand side. But for general purpose incubators, interestingly, the minimum fields appear at the top shelves. The reason is that the CO$_2$ incubator usually has a motor installed in the ceiling of the incubator, to circulate CO$_2$ and air. For general purpose incubators, without a strong CO$_2$ and air circulation motor, the top becomes the location furthest from electrical components, and therefore forms a minimum field. Another common characteristic of minimum fields is that they often appear at the left hand side of incubators. The reason is that most incubators are designed with their control panel located on the right hand side. Consequently, the electrical components tend to be installed on the right hand side, which allows the minimum field to occur on the left hand side.

### 3.2.2.4 Location of Maximum MF

Figure 15 shows the location of maximum MFs for different types of incubators.

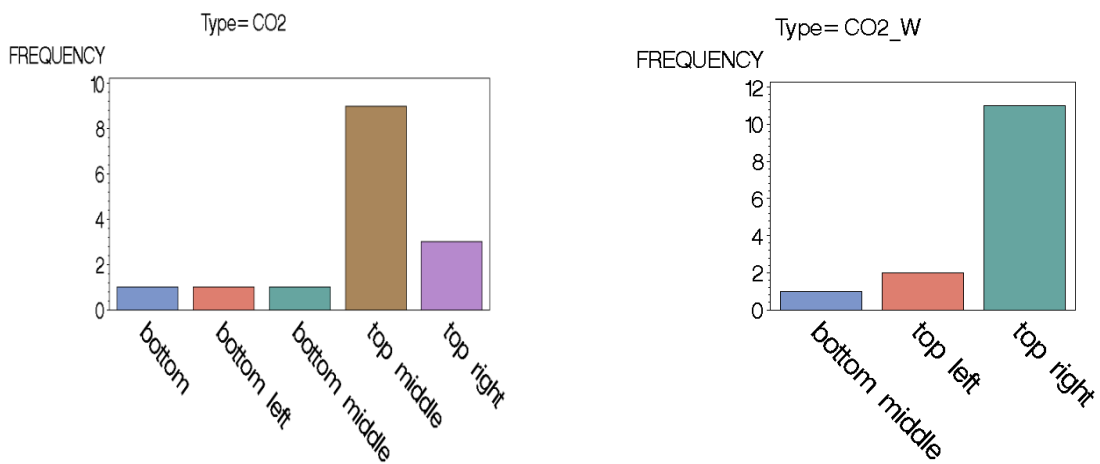

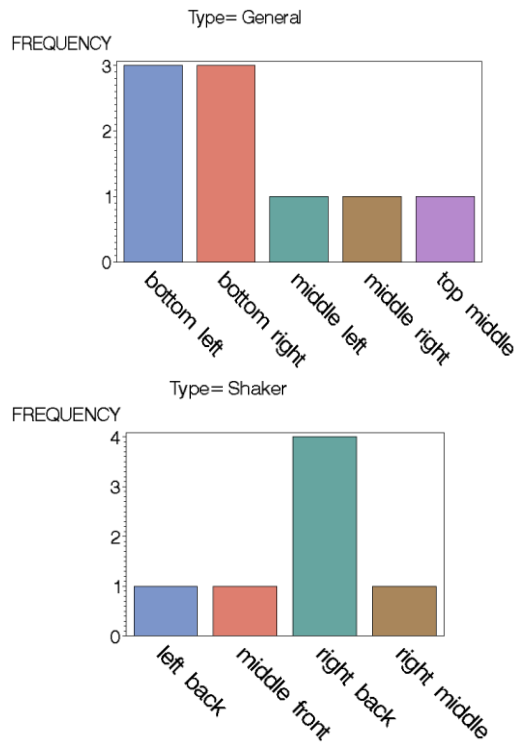

*Figure 15.* Location of Maximum MFs by incubator type.
Top left: $CO_2$ incubators.
Top left: $CO_2$ incubators with water jacket.
Bottom left: General purpose incubators.
Bottom right: Incubator shakers.

The maximum field is usually located at the opposite diagonal side of the minimum field. For the $CO_2$ incubator, this means that the top shelf to the right becomes a common maximum field location, and for the general purpose incubators, the bottom often becomes the place when MFs are highest. As for shaker incubators, the maximum fields more often appear at the right hand side, rather than the left hand side.

### 3.2.2.5 MFs over Frequency

A MF Frequency Spectrum for all 46 incubators was obtained at their centers, using the Narda EFA-300, which is capable of measuring from 5 Hz to 32 kHz. We focused on frequencies below 2 kHz. Although we have all the figures measured for every incubator, certain profiles come up very frequently. Spectra can generally be represented by the following two figures.

Class 1: Typical pattern of cell incubators (Figure 16). For most incubators (35 out of 46), the dominant frequency peaks are 60-Hz and 180-Hz. With only 2 exceptions, other peaks are at least 20 dB less than the 60-Hz level.

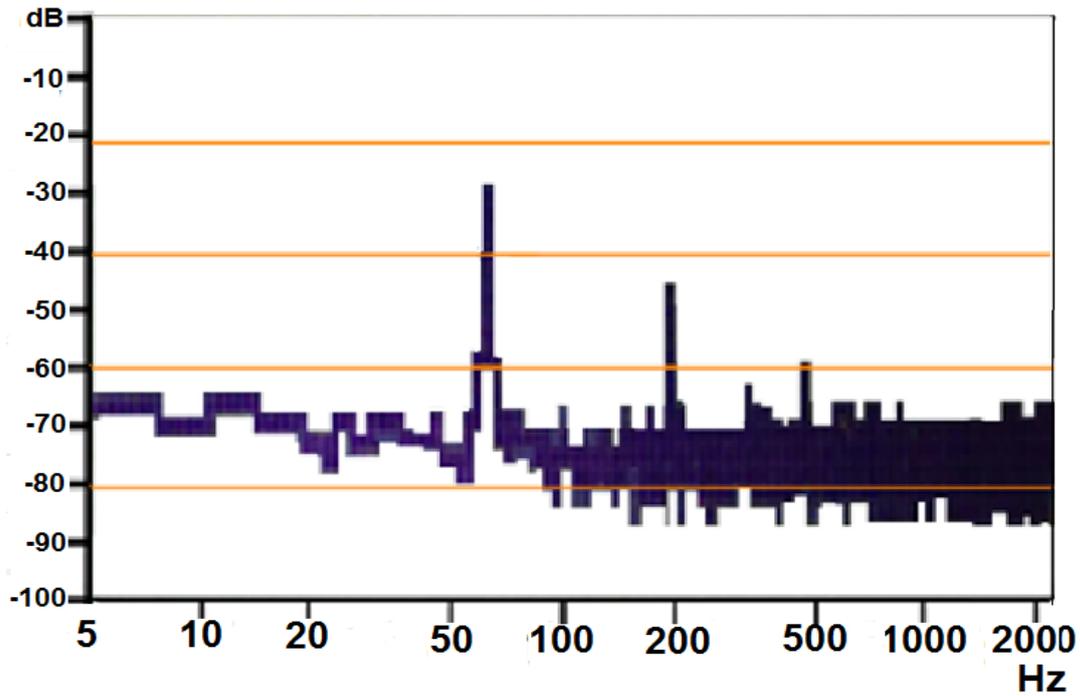

*Figure 16.*   *Class 1 Frequency Spectrum of Cell Incubators.*

Class 2: Older incubabors (11 out of 46) tend to have more complex frequency spectra, with more peaks, possibly resulting from older electrical systems that did not take advantage of micro-processors (Figure 17, from an old Revco incubator, model unrecognizable).

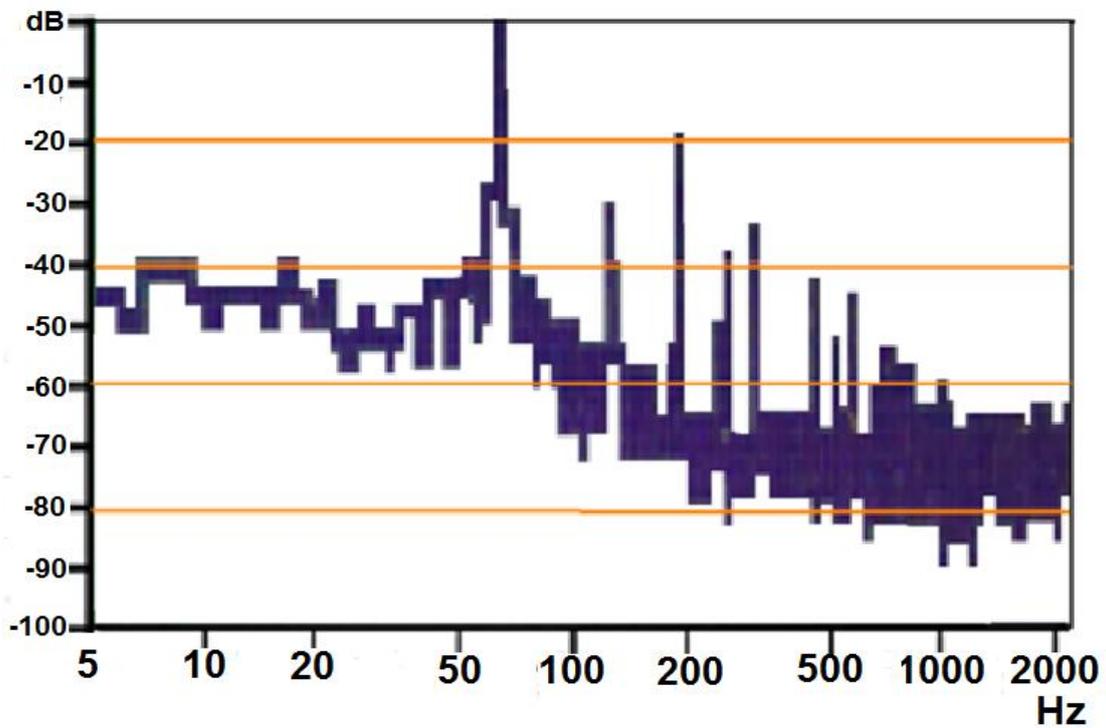

*Figure 17.*   *Class 2 Frequency Spectrum of older Cell Incubators.*

To sum up, although there are some frequencies other than 60-Hz inside incubators, they are generally much weaker (20 dB less) than 60-Hz. This means the field contributed by frequencies other than 60-Hz should be

$$P_1 = P_0\, 10^{\frac{L_{dB}}{10}}$$

$$P_1 = P_0\, 10^{\frac{-20}{10}} = 0.01\, P_0$$

where  $P_0$ :  energy related to 60-Hz
 $P_1$ :  energy related to the second peak, which is 20 dB less than

From the above calculation, we can see that the contribution of frequencies other than 60-Hz are negligible compared with 60-Hz, and should therefore be of sencondary concern.

### 3.2.2.6 MFs over Time

The Field Star 1000 measured 60-Hz MFs each second at the center points of all 46 incubators. The minimum recording time was 240 seconds.

MFs fluctuate over time, mainly due to the heater's on-off cycle. However, for $CO_2$ incubators, the $CO_2$ injection control relay and the electromagnetic valve's openings and closings also contributed to the fluctuations. Surprisingly, the patterns of the fluctuations are highly variable. Some of typical results are posted in Figure 18.

It should be noted that some studies (Berman et al. (1991), Lohmann et al. (2000), Guerkov et al. (2001), Selvamurugan et al. (2007), Schwartz et al.(2008), Sun et al. (2009 ) and Chen et al (2010)) reported that pulsed MFs have significant effects on various cells. Therefore, the changes in MFs which we observed, which are similar to pulsed fields, may be of concern.

Forma Scientific 3956

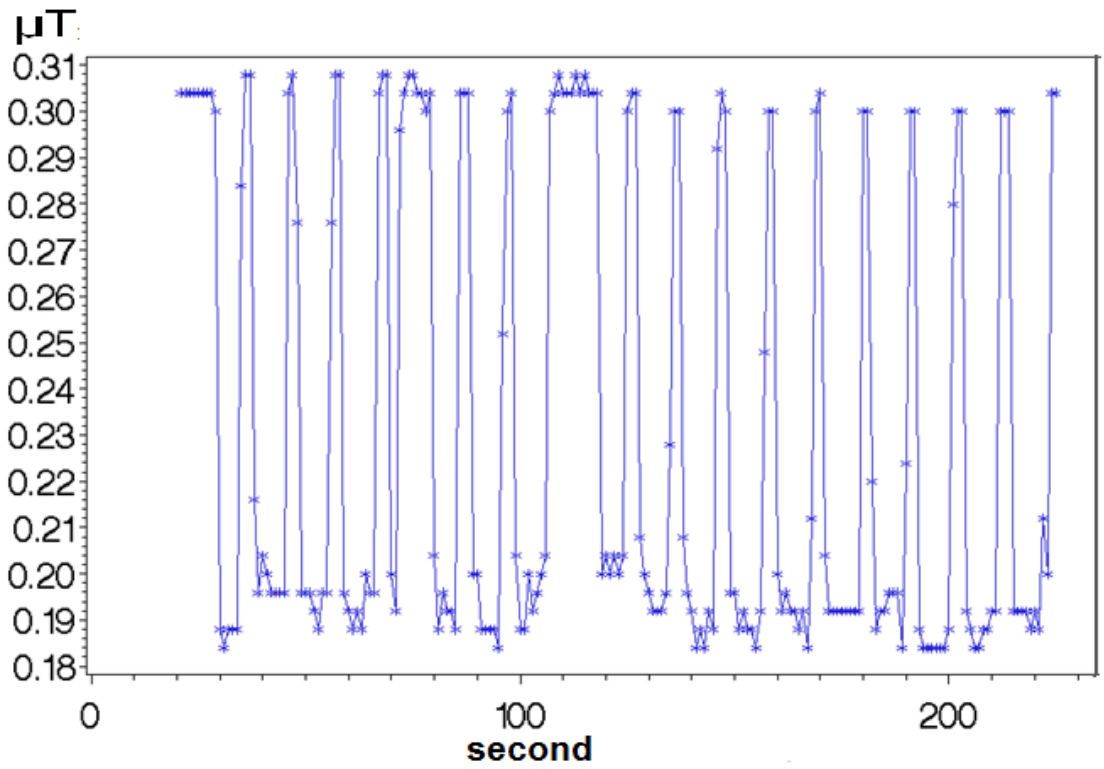

Nuaire US Auto Flow

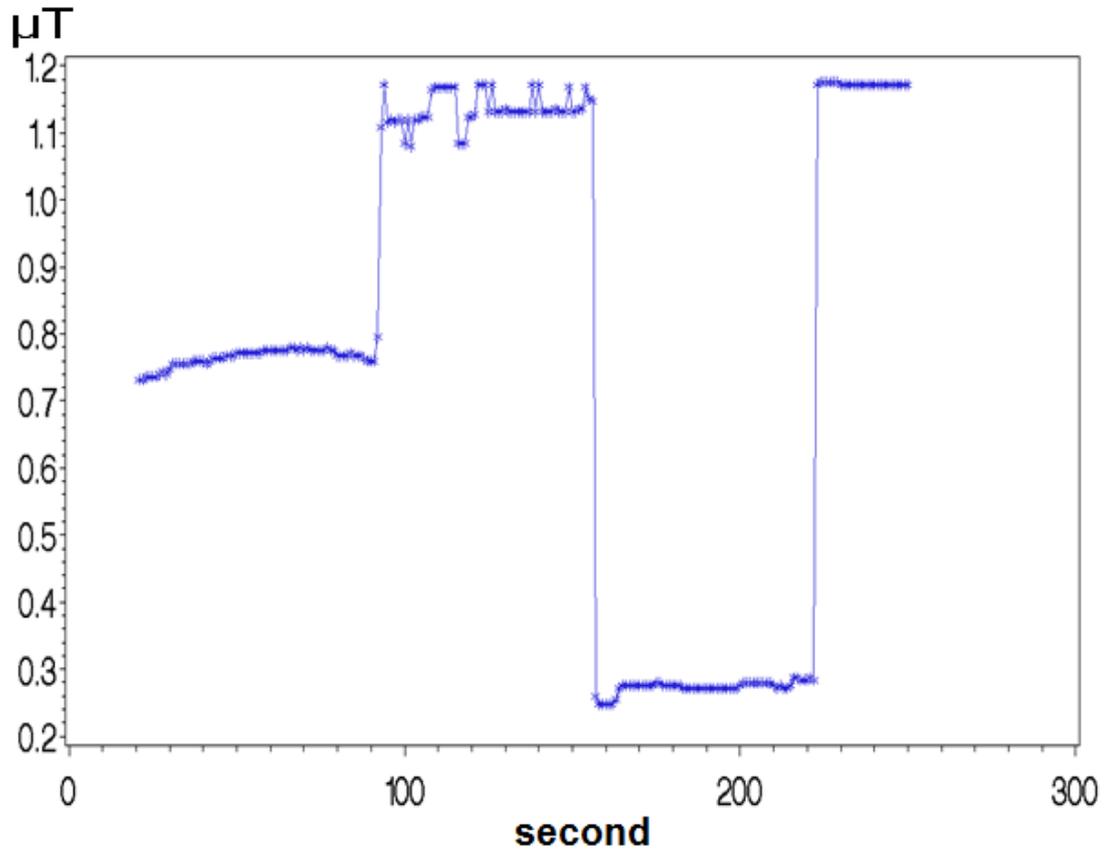

Forma Scientific 3110

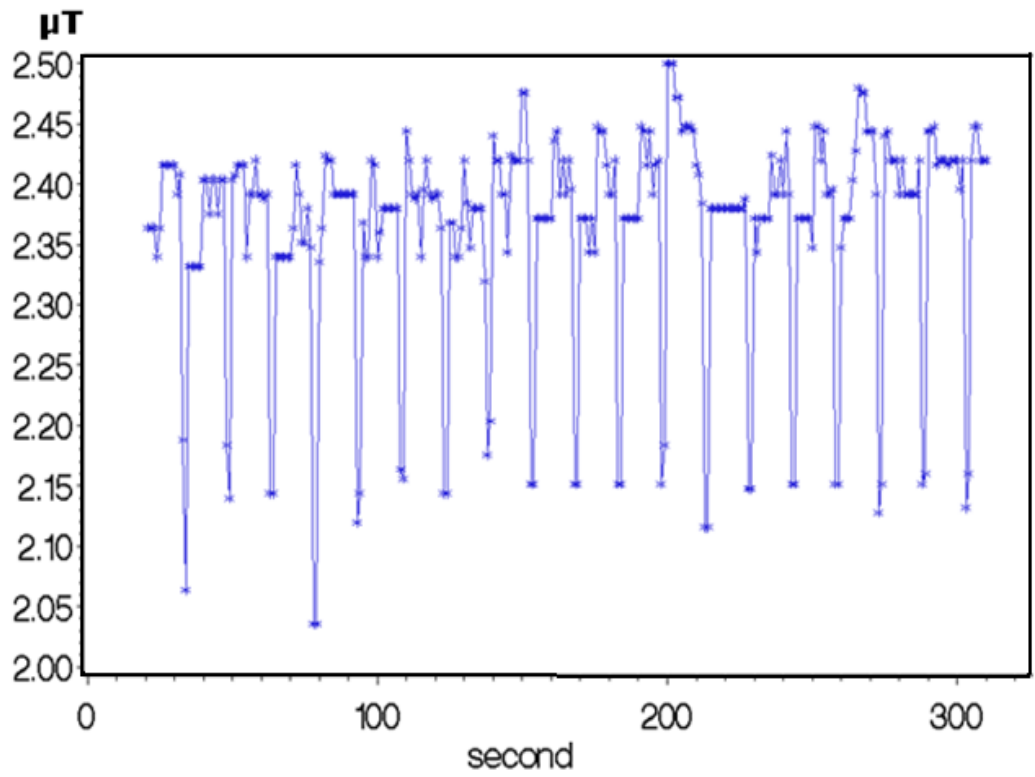

Gallenkamp (model N.A.)

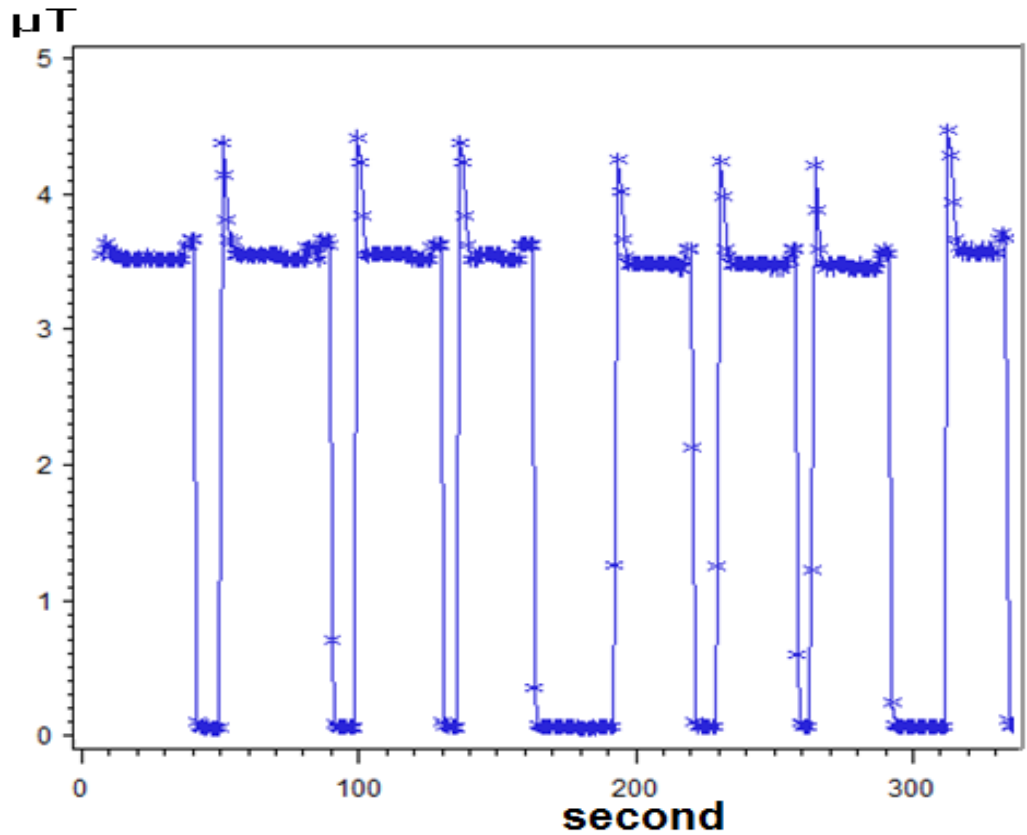

Fisher Scientific 546

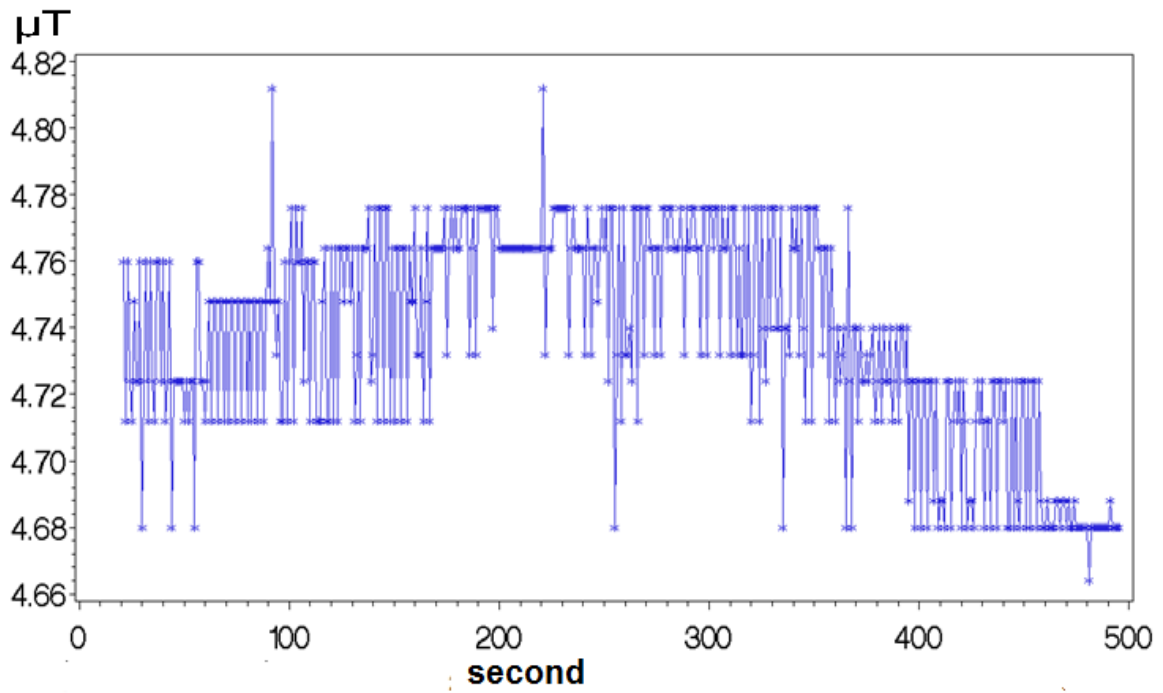

Napco 3550

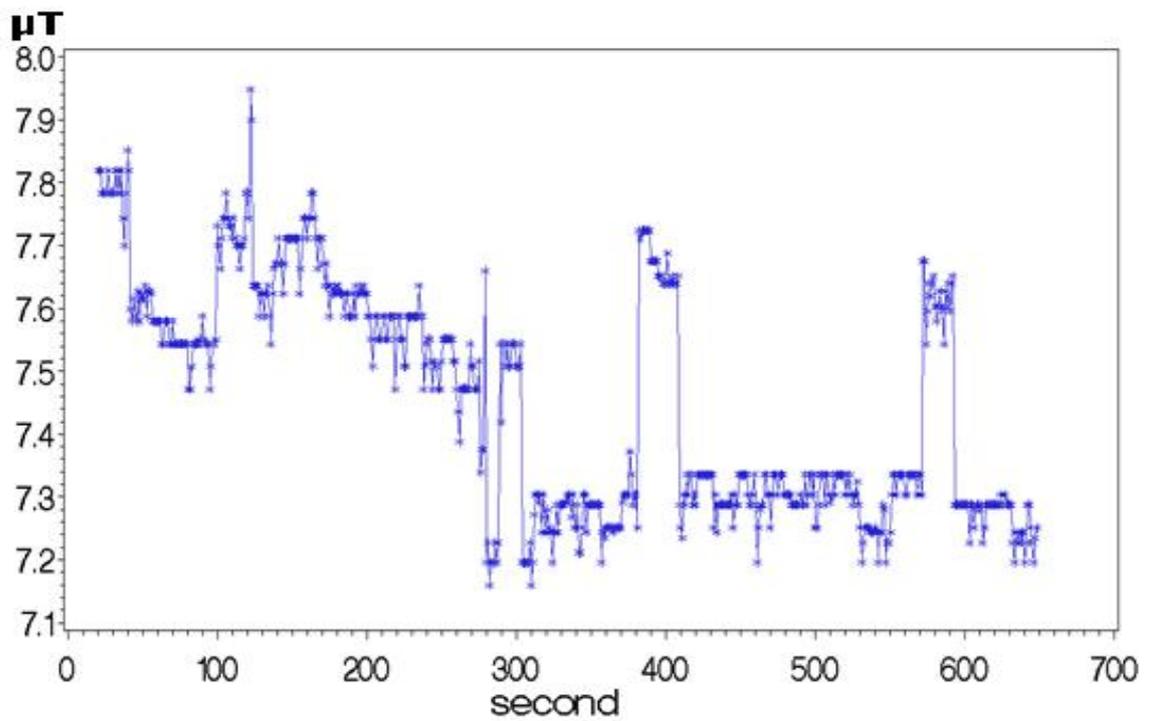

Fisher Scientific 610

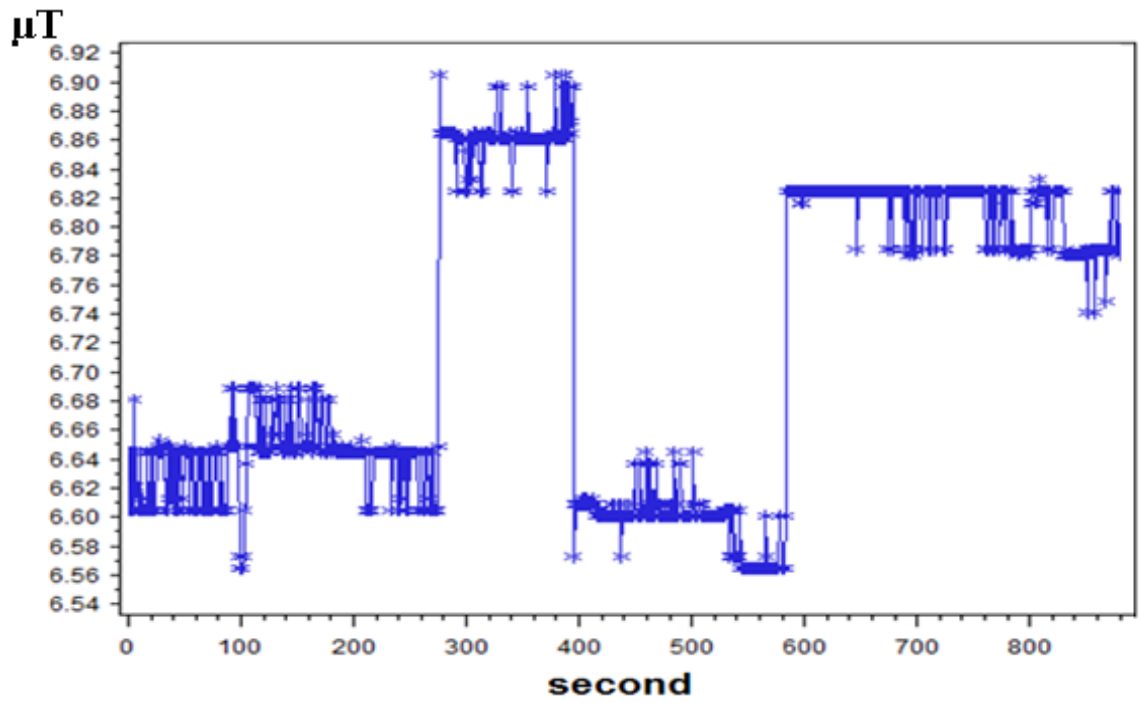

Fisher Sci 637D

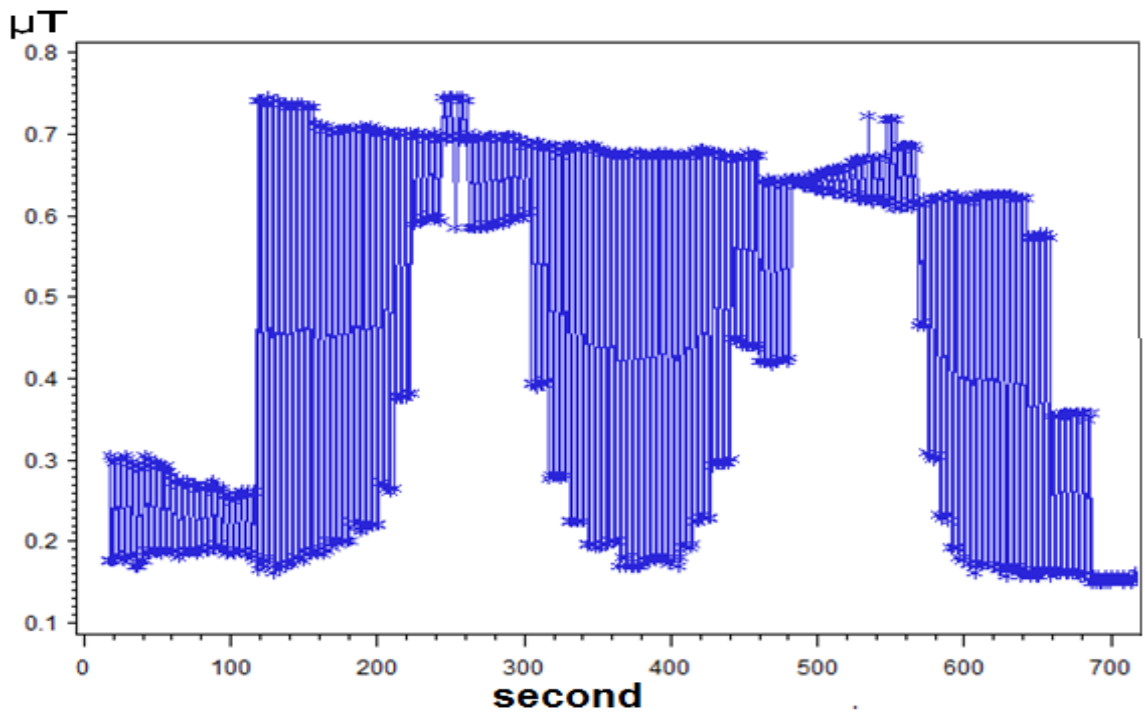

*Figure 18*.   Time patterns of 60-Hz MFs. Makes and Models are indicated at the top of each plot.

# 4 Discussion

## 4.1 Incubator MFs and precise *In Vitro* work

Among the 46 incubators surveyed, the lowest MF was 0.39 µT. According to Li and Héroux, transferring cells from extremely low fields (~ 0.004 µT) to 0.025 µT can cause cancer cells to lose chromosomes, and changes in the MF as small as to 0.01 µT can induce the same effect. Other observers, although they probably started from less well defined baseline MFs, reported effects on cells under MF exposures as low as 0.2 µT (Table 1), substantially less than 0.39 µT. Therefore, our survey results should serve as an alert to bioscience researchers and incubator manufacturers.

Li and Héroux have also reported effects of static MFs, but these effects became detectable only under very low levels of ELF MFs. This survey did not compile static MFs.

## 4.2 Users are Unaware

A second finding of this survey is that many incubator users are unaware of the existence of MFs inside their incubators, nor do they realize that MFs play a role in their experiments. We found that many incubator operators like to place their culture vessels on the top shelf, and even stack them high. This may reflect a belief that heating elements are placed at the bottom, coupled with ignorance of the high MFs associated with air circulation ventilators. According to our survey, the top shelf is usually the place where maximum MFs are found.

## 4.3 Manufacturers are Unaware

One may wonder if new incubators have weaker fields than older ones, reflecting a positive evolution in equipment design. Unfortunately, we cannot answer this question formally, because during our survey, the information about the year of make of incubators was often inaccessible. But our impression, from correlating observed field values with apparent equipment age, is that incubator engineering is, today, blind to the requirements of serious environmental controls. Rather manufacturers aim at providing more electrically-based conveniences, increasing the complexity of the electro-magnetic incubator environment. Consequently, we draw the conclusion that incubator manufacturers are not making progress in minimizing or controlling the MFs of their products.

## 4.4 Control Measures for Users

What should the conscientious operator do, in the absence of re-designed equipment? Our suggestions are as follows.

A.  When placing cell cultures in incubators, choose the lowest field location within the unit, and use that location repeatedly for a coherent experimental series. This particular control measure is already part of the routine of many experienced cell culture technicians. Avoid using the maximum field regions. The most prominent effect of MFs on cancer cells (Karyotype Contraction) appears to be related to the time-weighted-average of the MF (Li and Héroux, 2012).

B.  Since incubator walls only weakly attenuate external fields, choose the location of an incubator carefully, so that the MF environment is as weak and constant as possible (preferably below 0.1 µT, i.e. half of 0.2µT). Cells in a constant MF seem to adapt to a fixed value over time (Li and Héroux, 2012).

B.  Do not place two or more incubators together without a minimum clearance of 20 cm, to avoid cross-incubator interference.

C.  Insure that no occasional source of MFs is activated or brought in proximity to the incubator during culture growth.

## 4.5   Control Measures for Manufacturers

Abating MFs inside incubators through shielding or active MF compensation would be expensive. However, simple controls of the various MF sources can be economically implemented .

Taking as an example the door of the Baxter WJ501 incubator (Figure 19), which we previously assessed, we can see the heating wire widely spread over the door surface. If the heating wire is folded back upon itself (and the pair twisted), the MF can be greatly reduced.

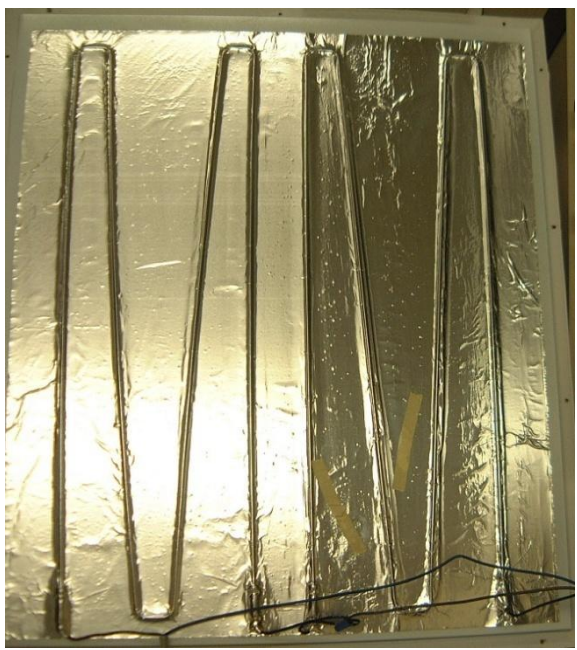

***Figure 19**.   The heating wire inside the door of the Baxter WJ501 incubator.*

Another simple way to minimize MFs is to rearrange the layout of electrical components. Taking again the Baxter incubator as an example, we found that a circuit board with a transformer was installed at the center, vertically, and towards the front of the unit (Figure 20). This created a MF peak on the opposite side of the incubator wall, where users are very likely to lay their cultures. We suggest that the manufacturer move those parts that are major sources of MFs, such as motors and transformers, to locations users are least likely to use, for example, the top corners, and clearly mark or even obstruct internally those maximum MF regions inside the incubator.

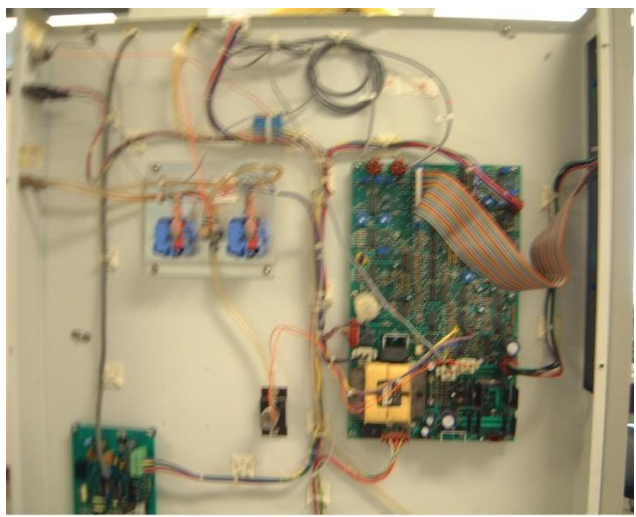

**Figure 20**. *Photo of electrical components laid on the left hand side of the Baxter WJ501 incubator.*

There is still uncertainty today as to whether MF effects are due to the MF itself exclusively, or to induced currents. If induced currents are involved in some biological effects, then reduction of the time variations of MFs is an objective. According to the Maxwell–Faraday equation,

$$\oint_{\partial S} \mathbf{E} \cdot \mathrm{d}\mathbf{l} = -\frac{\partial \Phi_{B,S}}{\partial t}$$

higher rates of MF change over time generate a higher electric field, which means stronger currents introduced inside electrolytes. Since some studies found that pulsed MFs (with a high rate of MF change over time) have significant biological effects, restricting the electromagnetic bandwidth to low values should be a goal for incubator designers. This can be implemented by using DC rather than AC heating, with a proportional rather than discrete power control to maintain temperature levels. Switching relays for heating should be replaced by proportional amplifiers. Static field levels can be limited by using proper return paths for heating currents, and should be smaller than background earth levels (50 µT). The electro-magnetic induction of the transient MFs generated by the operation of solenoid valves that control $CO_2$ gas flow could be reduced by providing a separate switching unit that can be remotely situated, or by solenoids shielding and filters.

If manufacturers were really serious about delivering top-level instruments that insure reliable culture conditions, they could include in their equipment simple electromagnetic sensors that would warn users if electromagnetic fields reached troublesome levels. This could be built using tri-axial magnetic sensors coupled to analog or digital vector integration feeding a threshold detection circuit, and finally an alarm.

# 5 Conclusion

All incubators surveyed were found to have average ELF MFs higher than 0.39 µT, high variability within their useable space, as well as high variability over time. Since there is evidence that ELF MFs influence cell cultures and other experimental models, and since we have been able to demonstrate that currently used incubators expose their contents to significant levels of such fields, manufacturers of incubators, and researchers using such devices should endeavor to control ELF MFs to the lowest levels achievable in their experiments in order to prevent their influence on scientific results.

# 6 Acknowledgements

We thank Michel Bourdages and Duc Hai of Institut de Recherche d'Hydro-Québec for supplying the Narda EFA-300 for the purposes of this survey. We thank Dr. Ying Li for reviewing the document.

# Appendices

**A   Plots of X-axis MF (Baxter WJ501)**
**B   Plots of Y-axis MF (Baxter WJ501)**
**C   Plots of z-axis MF (Baxter WJ501)**
**D   Plots of integrated MF (Baxter WJ501)**
**E   Plots of MF produced from the door heater of Baxter WJ501**
**F   Plots of integrated MF (heater off, Baxter WJ501 )**
**G   Plots of MF of Forma 3310 (heater on)**

**Appendix A: X-axis (side to side) MFs at different heights for Baxter WJ501**

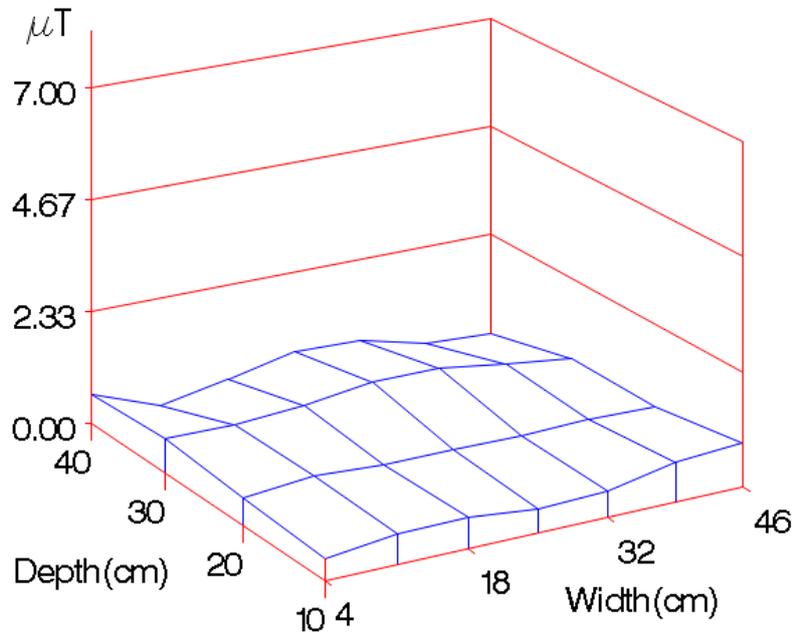

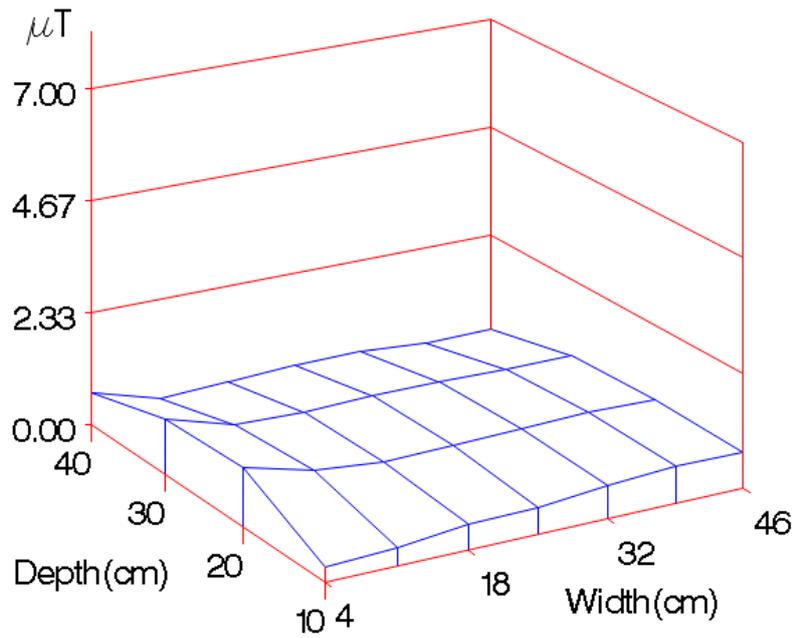

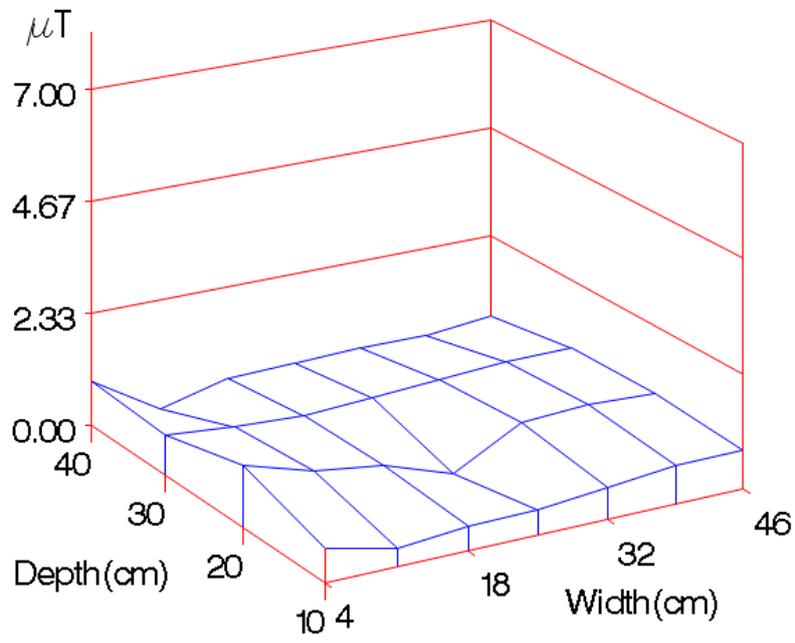

Magnetic Field In Direction X At Height= 14.6cm

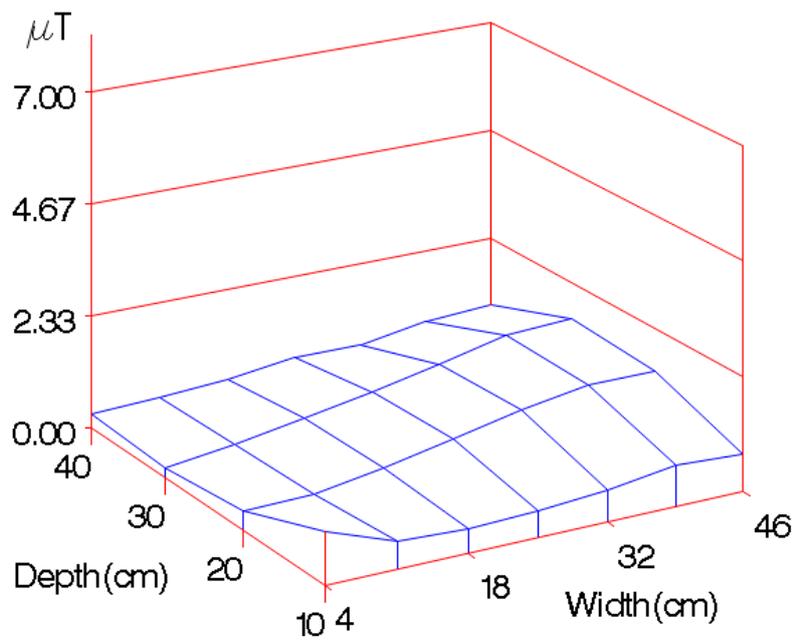

Magnetic Field In Direction X At Height= 21.9cm

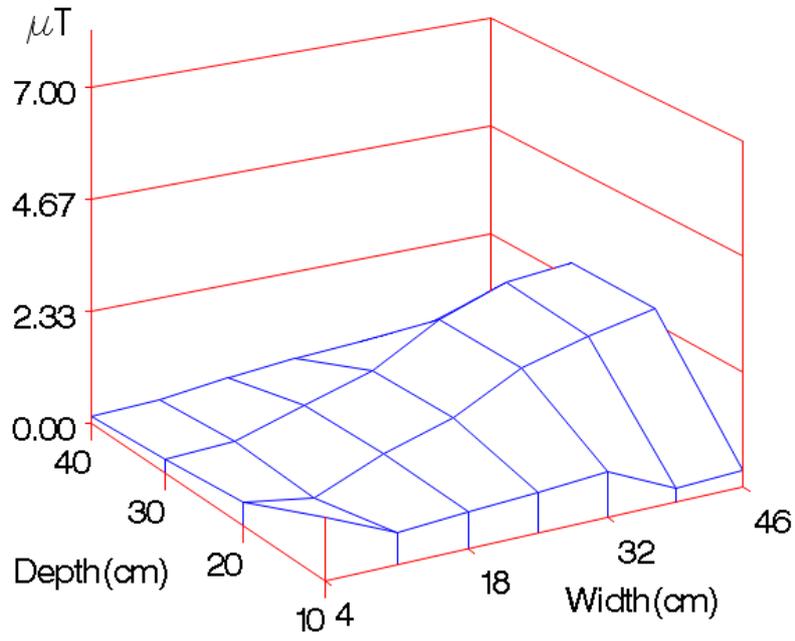

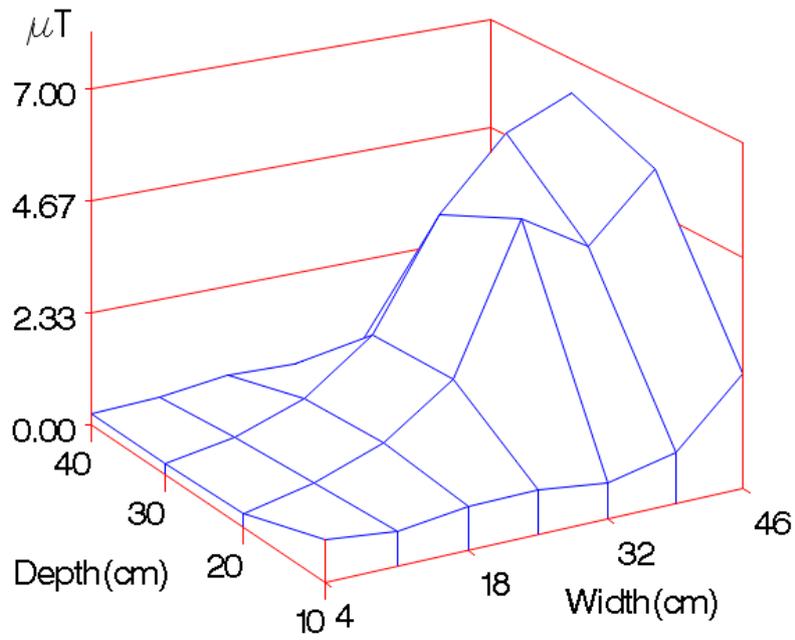

**Appendix B: Y-axis (front to back) MFs at different heights for Baxter WJ501**

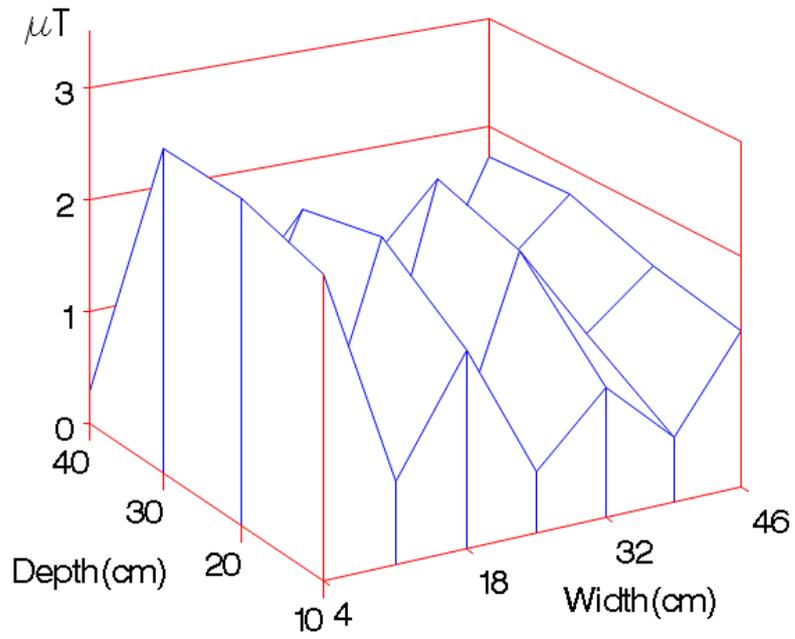

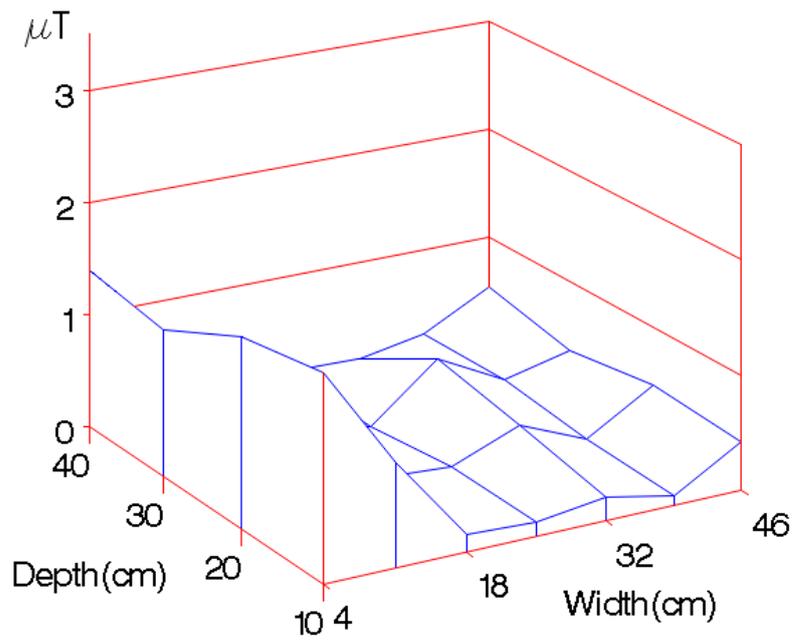

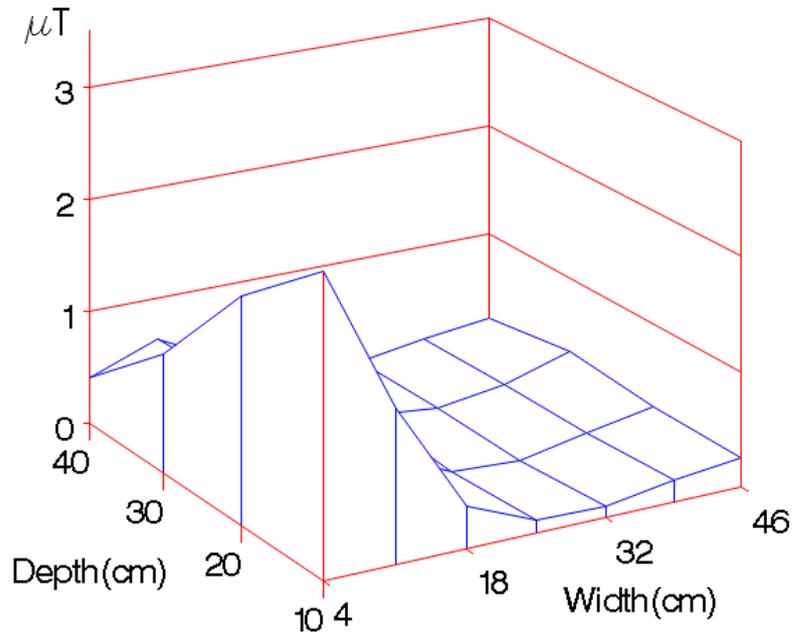

Magnetic Field In Direction Y At Height= 14.6cm

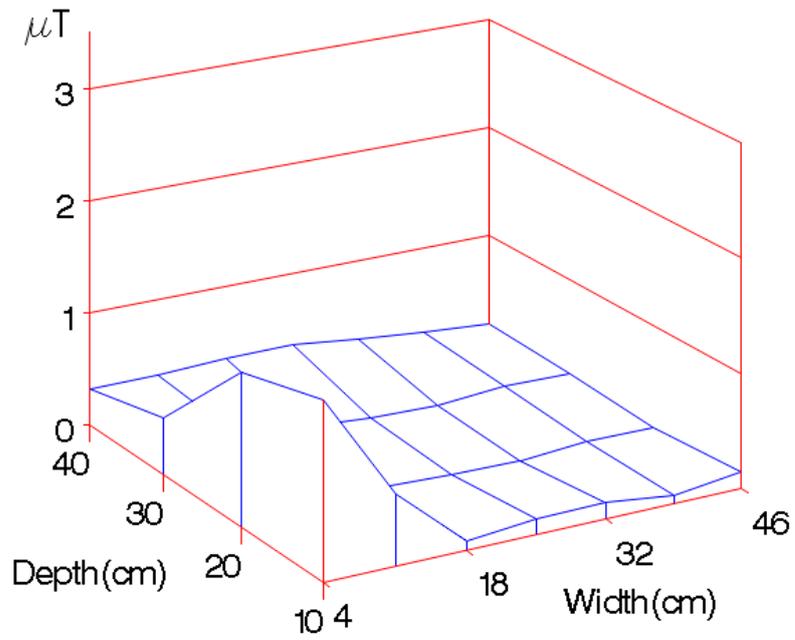

Magnetic Field In Direction Y At Height= 21.9cm

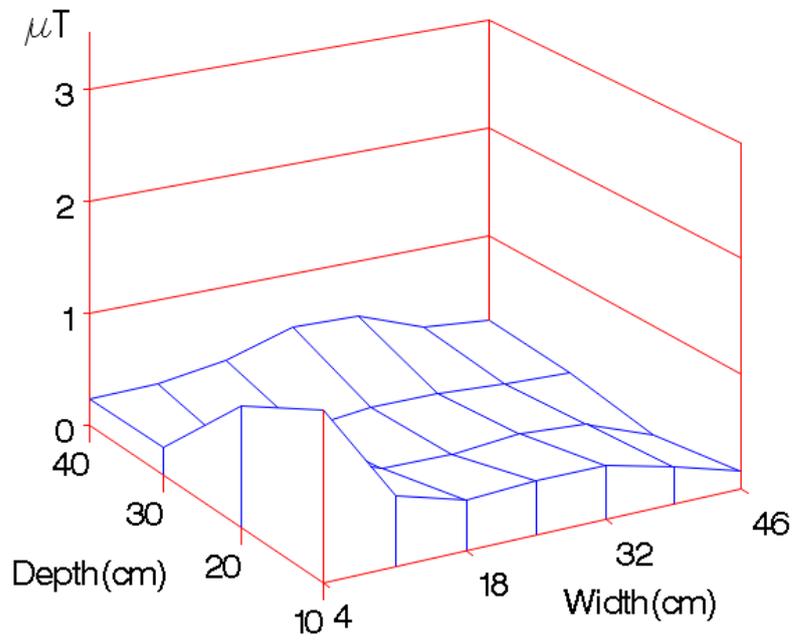

Magnetic Field In Direction Y At Height= 29.2cm

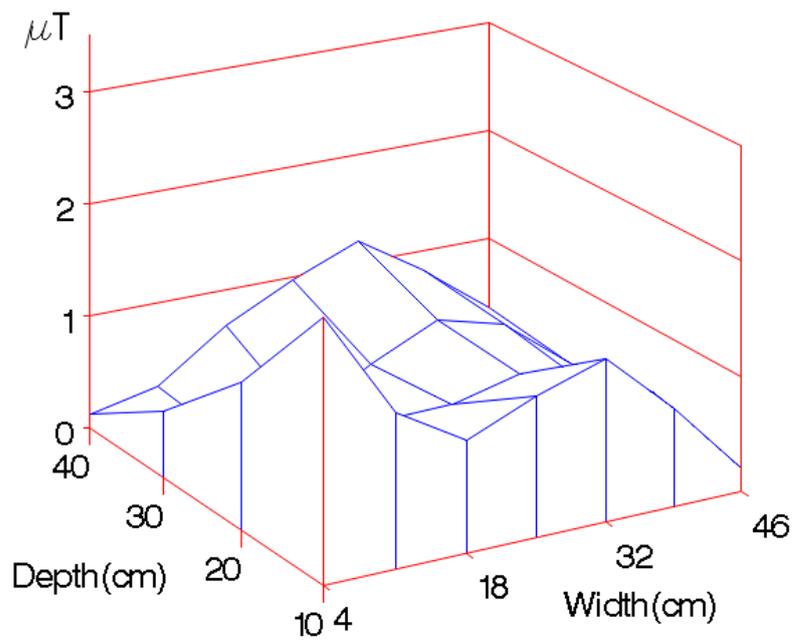

Magnetic Field In Direction Y At Height= 36.5cm

**Appendix C: Z-axis (bottom to top) MFs at different heights for Baxter WJ501**

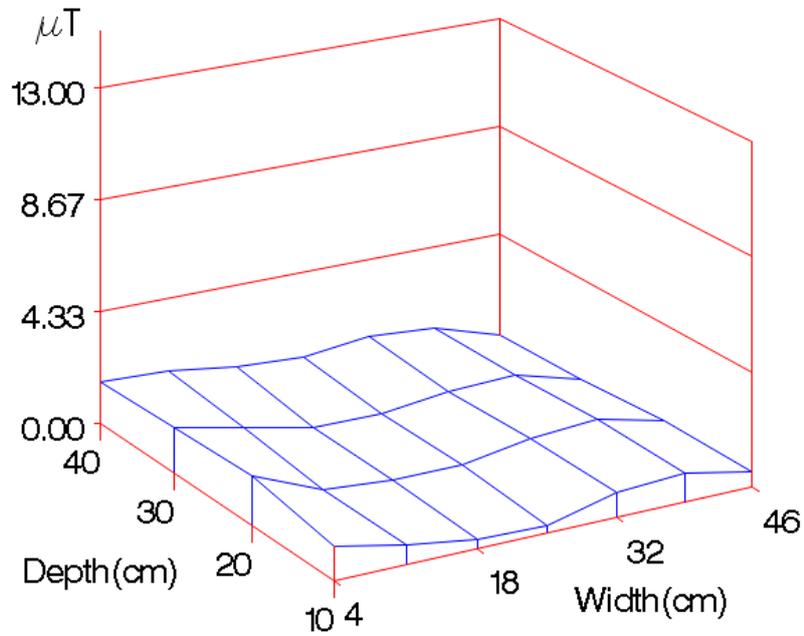

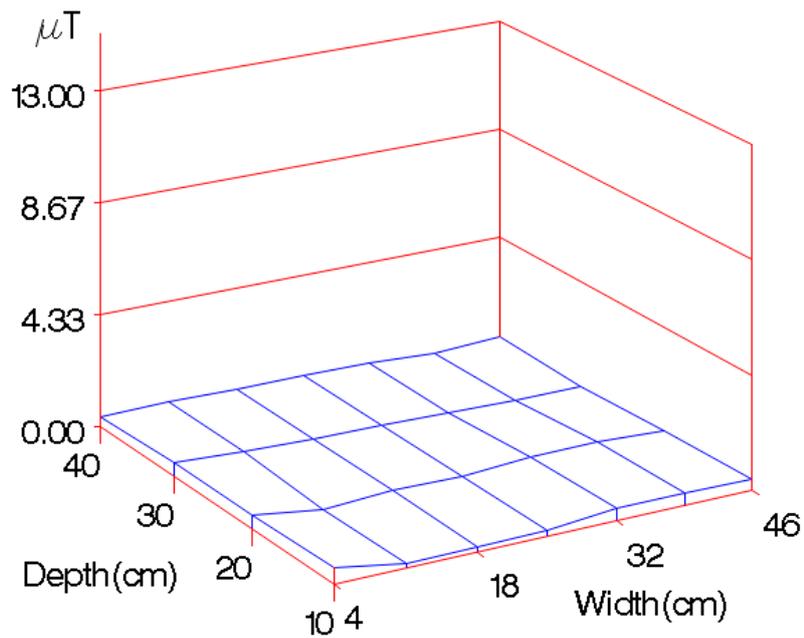

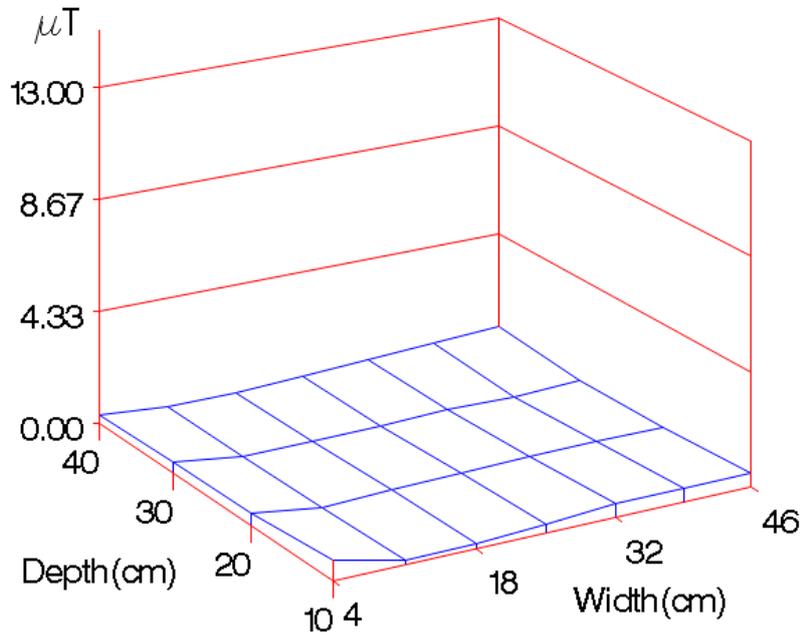

Magnetic Field In Direction Z At Height= 14.6cm

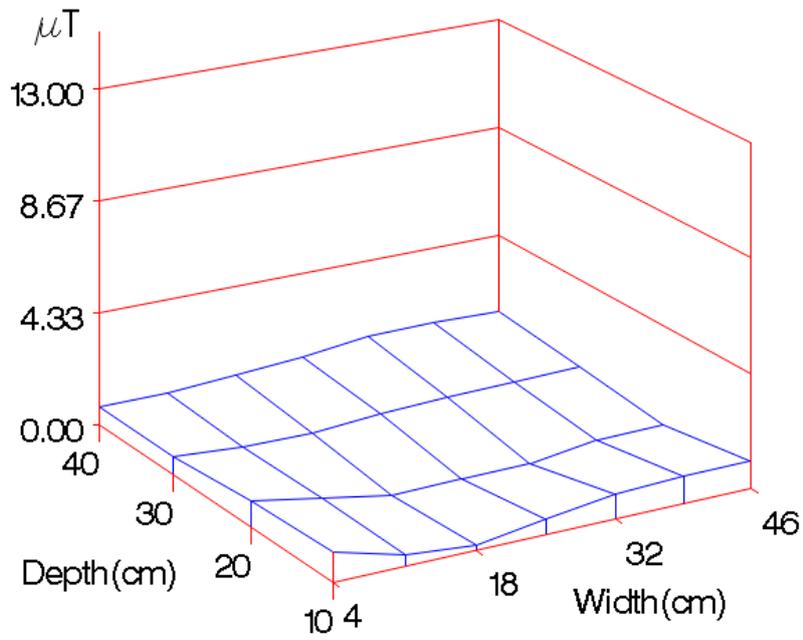

Magnetic Field In Direction Z At Height= 21.9cm

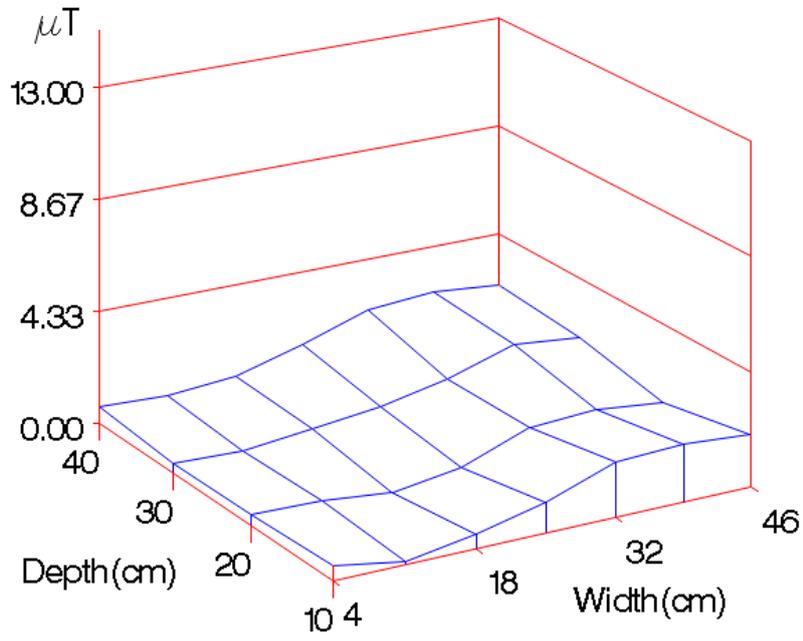

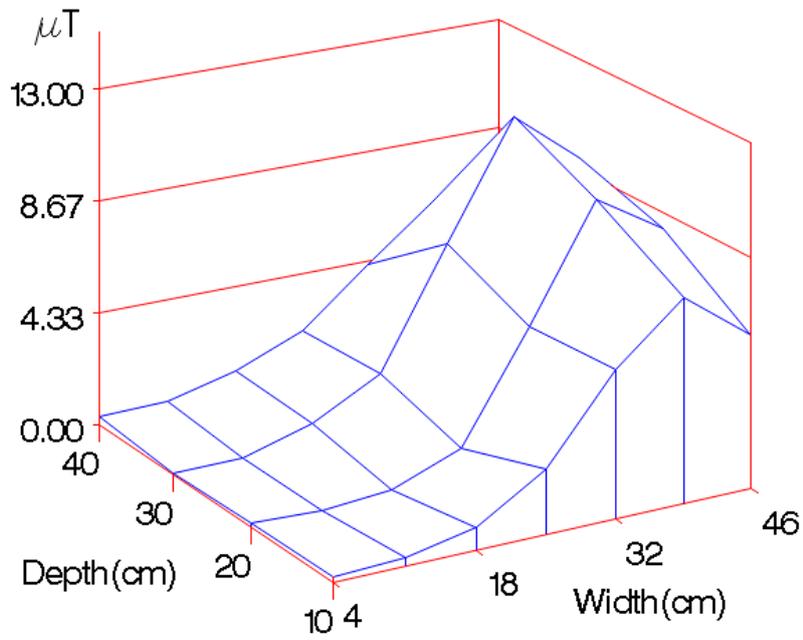

**Appendix D: Integrated MFs (X-Y-Z) for Baxter WJ501**

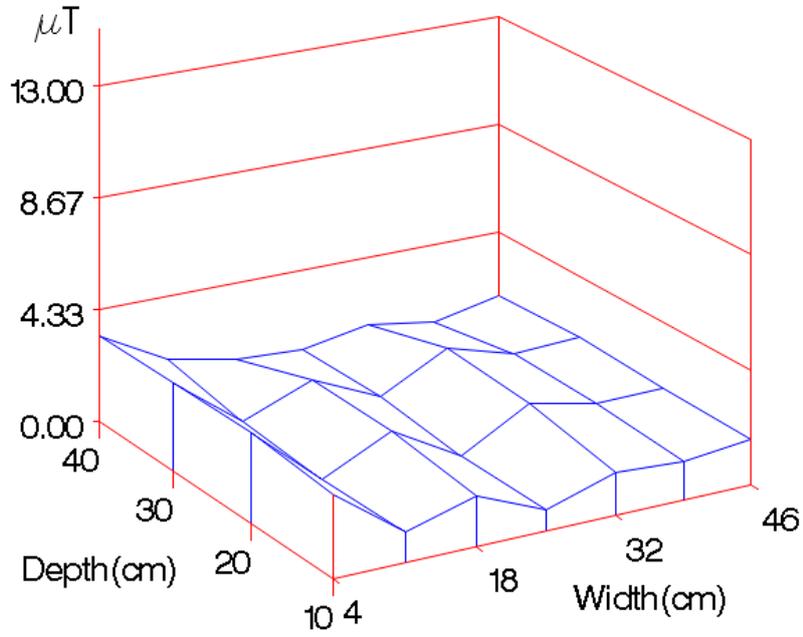

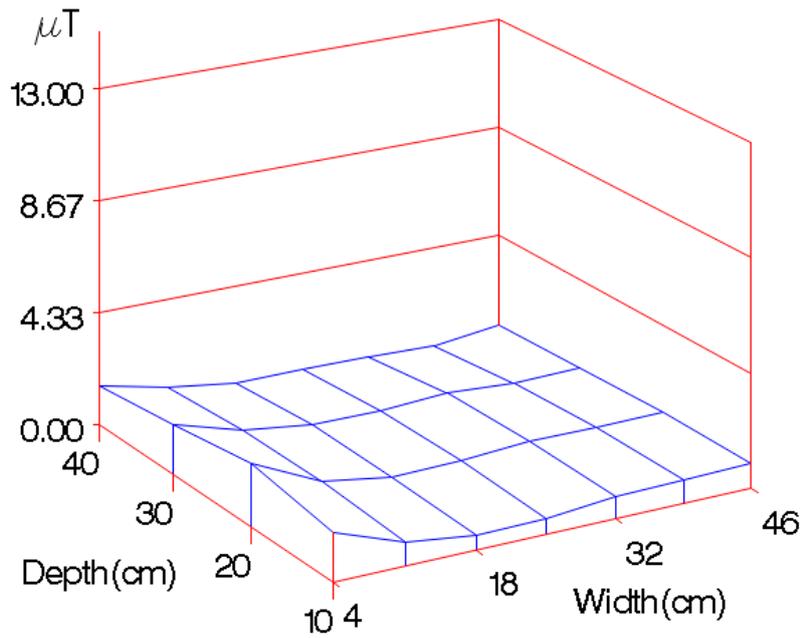

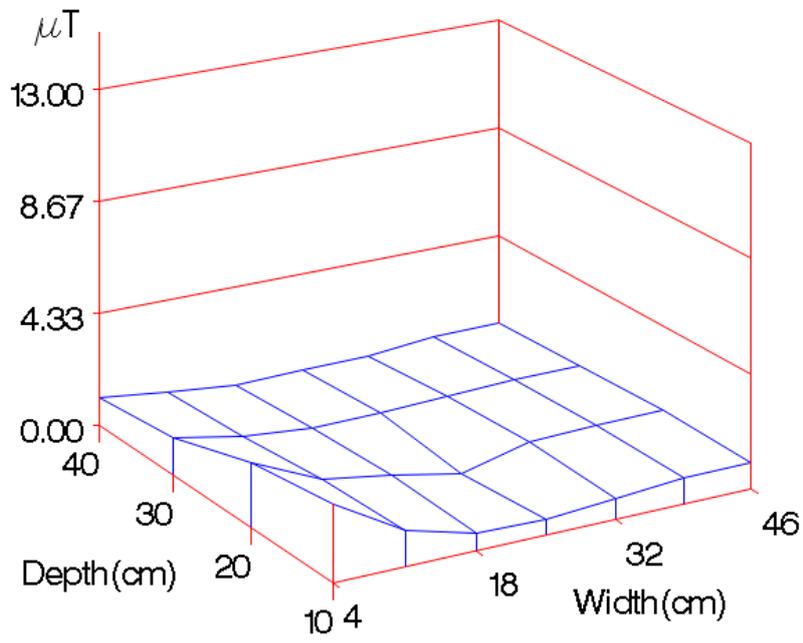

Integrated Magnetic Field At Height = 14.6cm

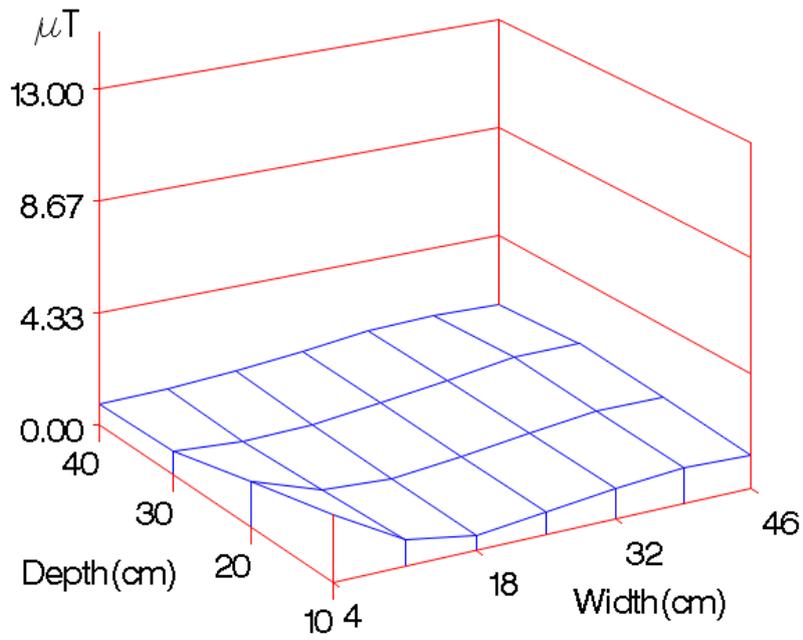

Integrated Magnetic Field At Height = 21.9cm

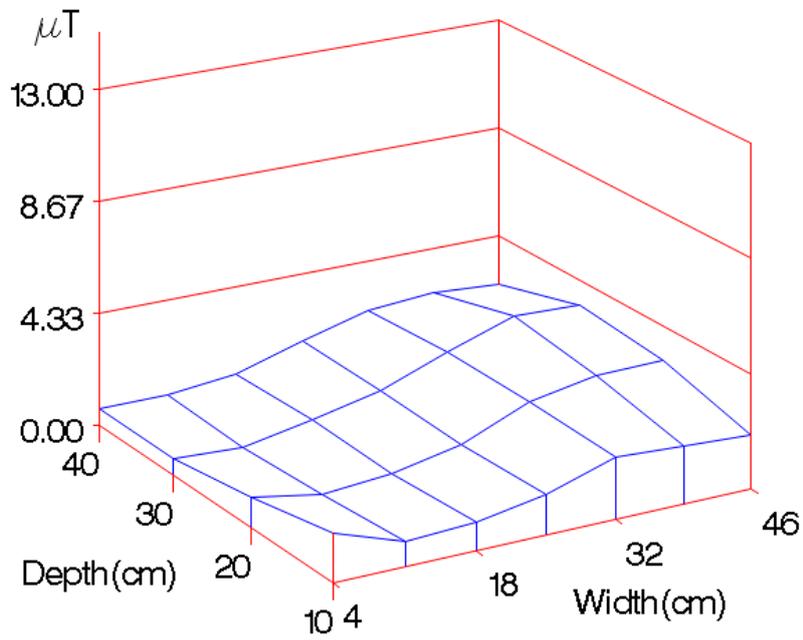

Integrated Magnetic Field At Height = 29.2cm

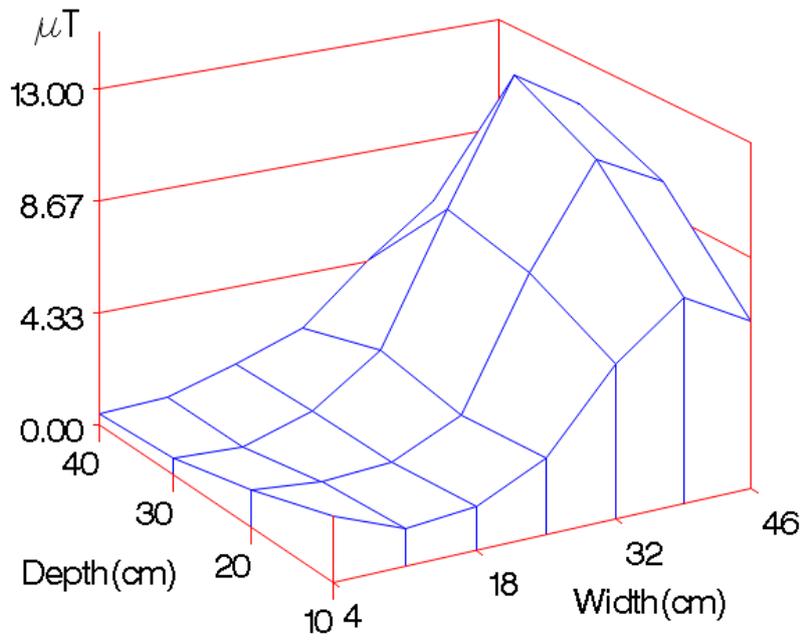

Integrated Magnetic Field At Height = 36.5cm

**Appendix E: Integrated MFs (X-Y-Z) from maximum door heating for Baxter WJ501**

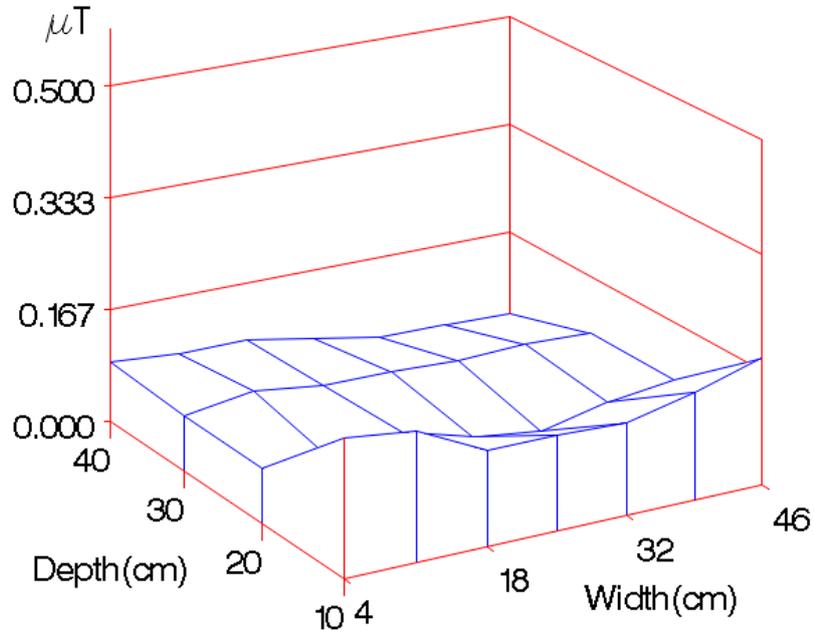

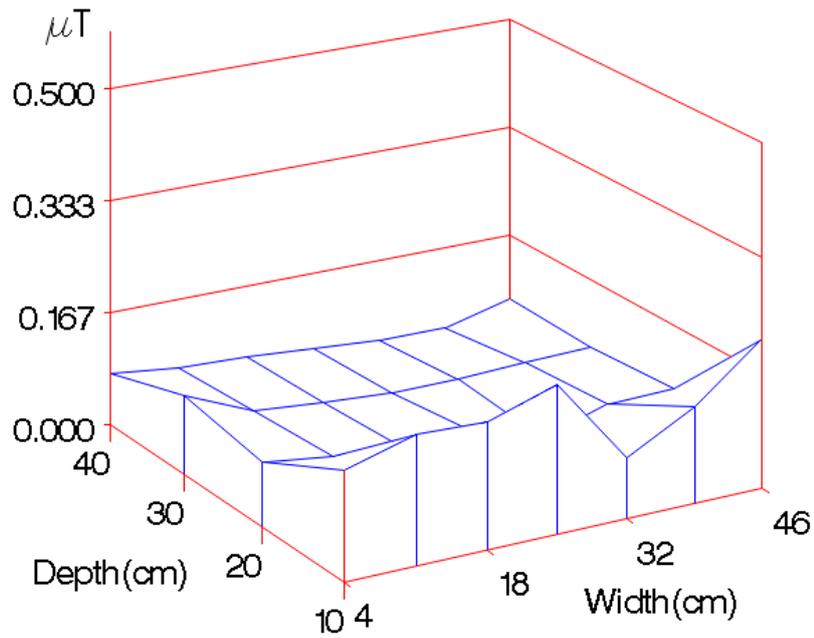

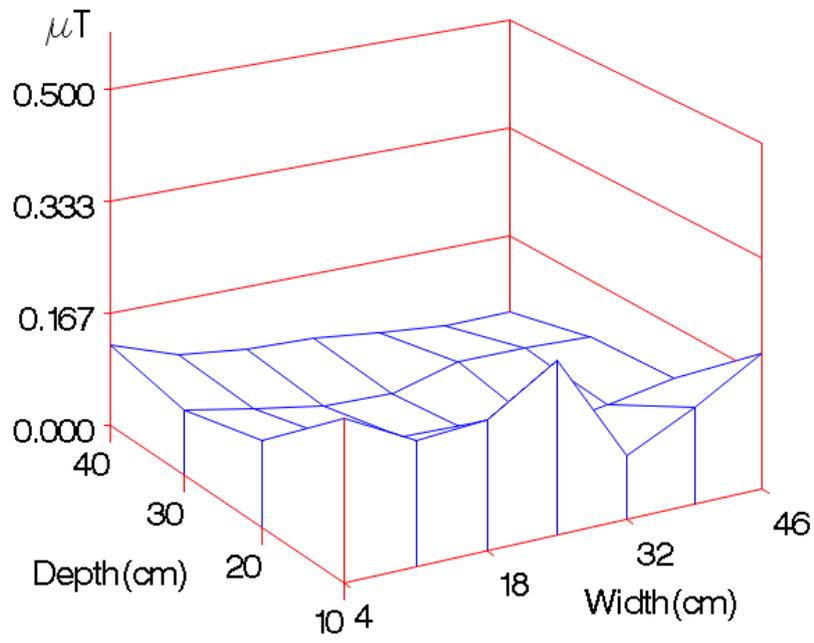

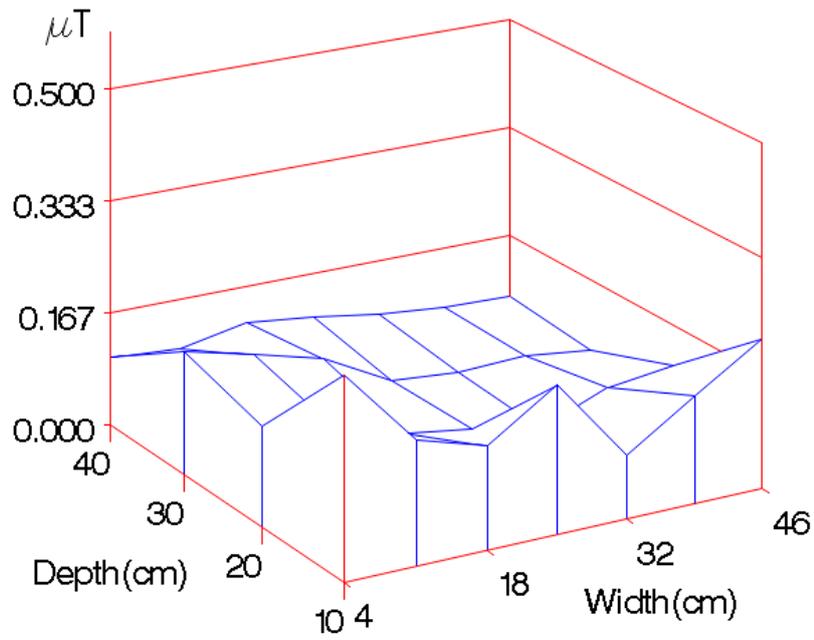

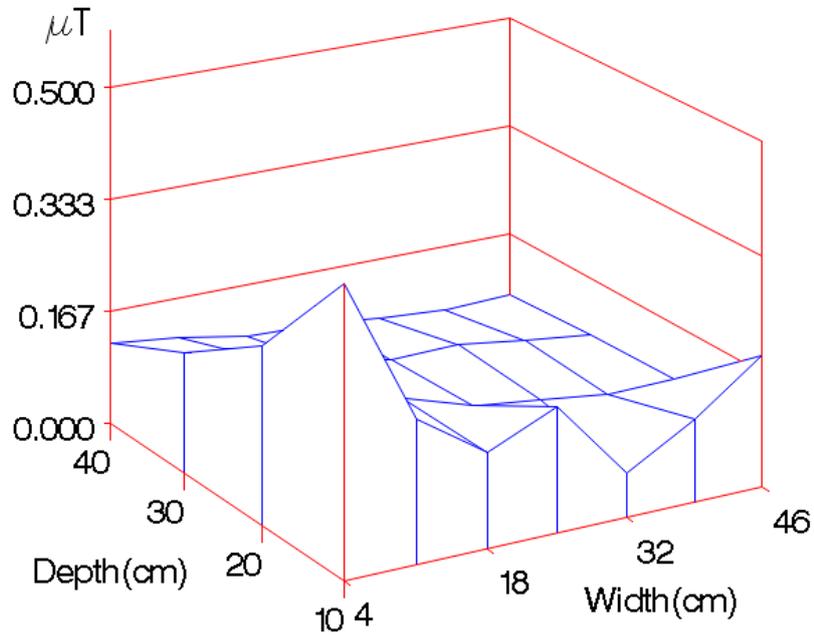

Integrated Magnetic Field At Height = 29.2cm

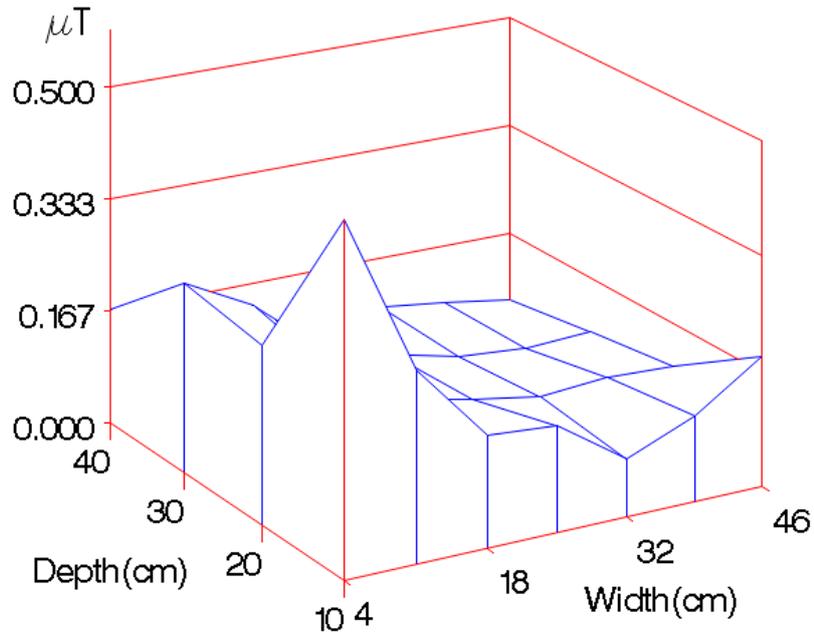

Integrated Magnetic Field At Height = 36.5cm

## Appendix F: Integrated MFs (X-Y-Z) for Baxter WJ501 with heater "off"

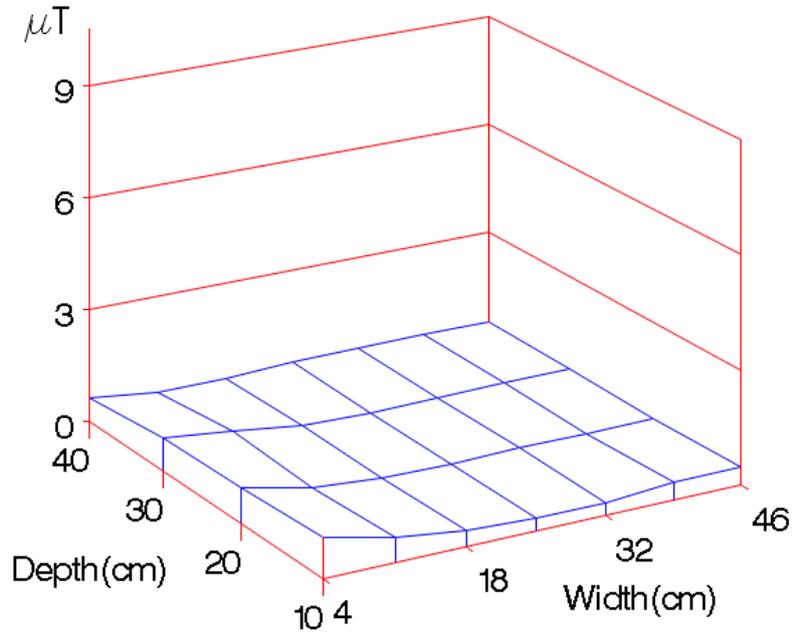

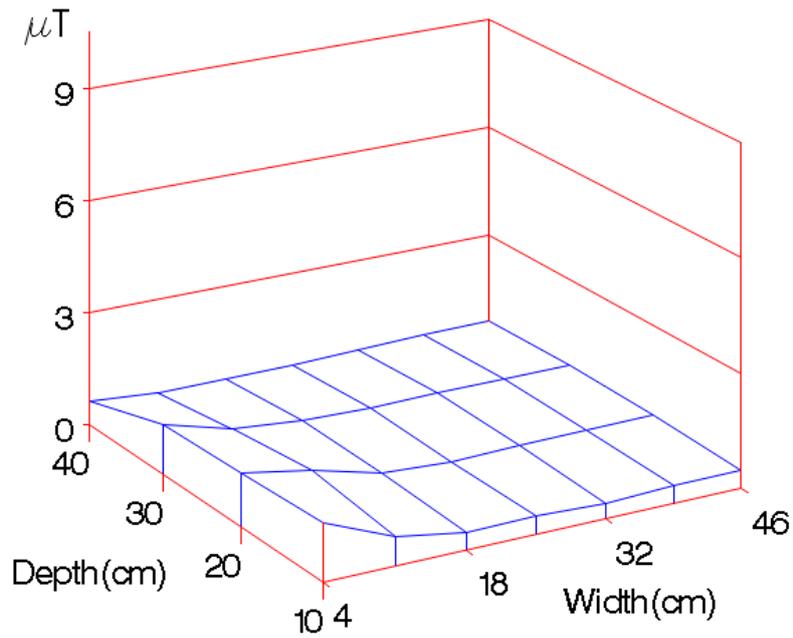

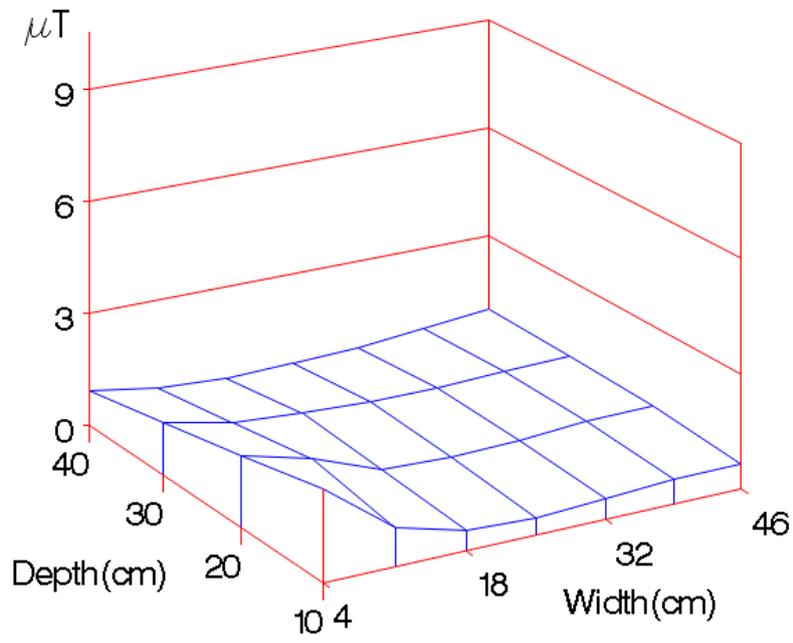

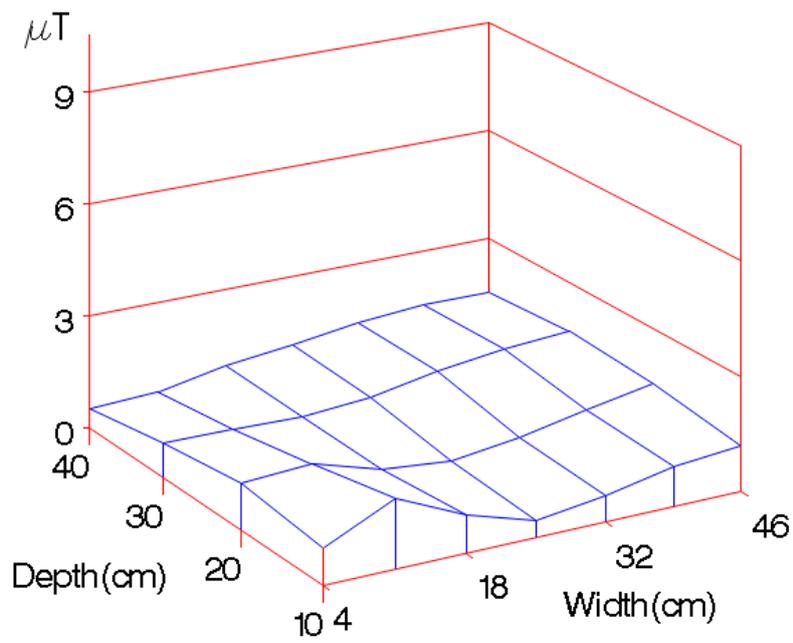

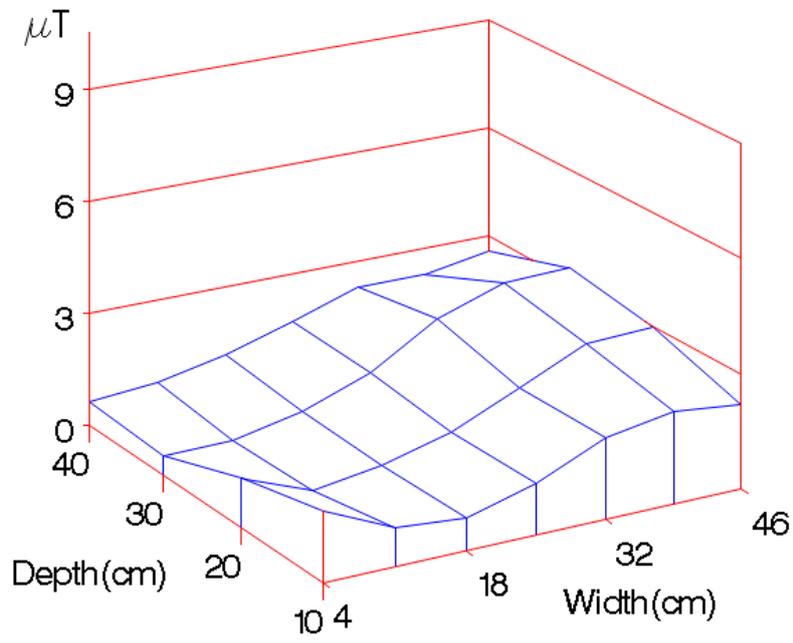

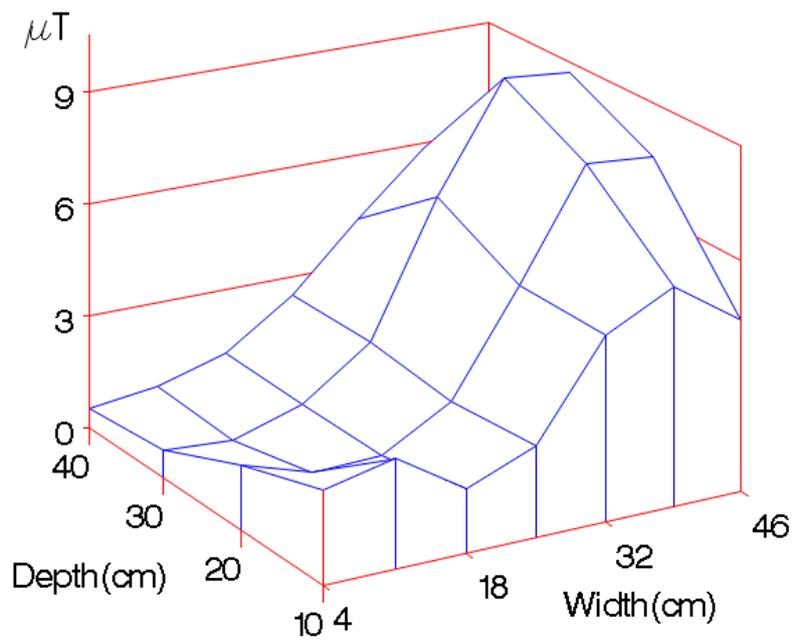

## Appendix G: Integrated MFs (X-Y-Z) for Forma 3310 with heater "on"

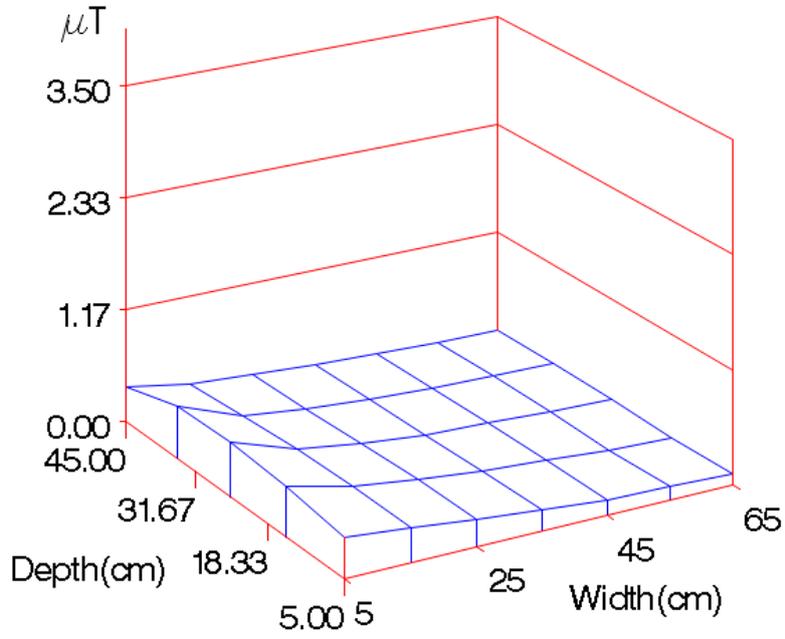

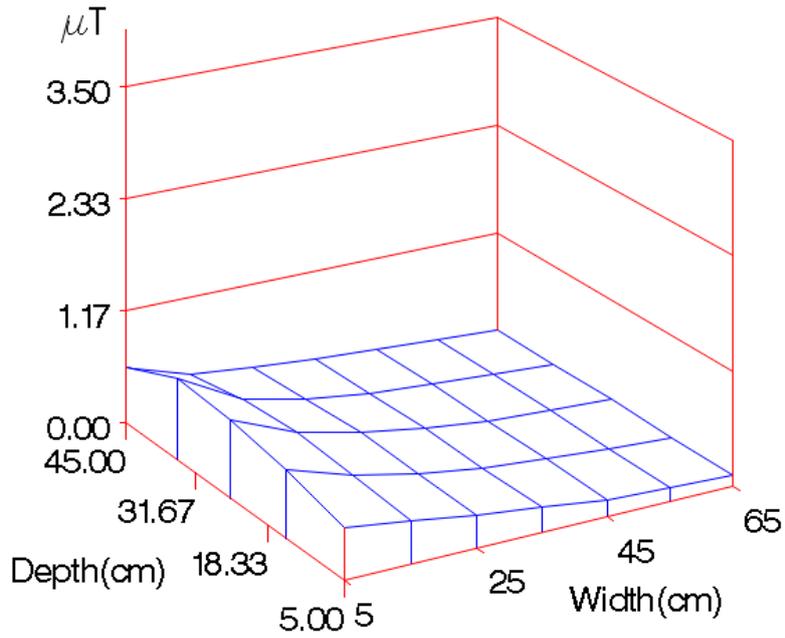

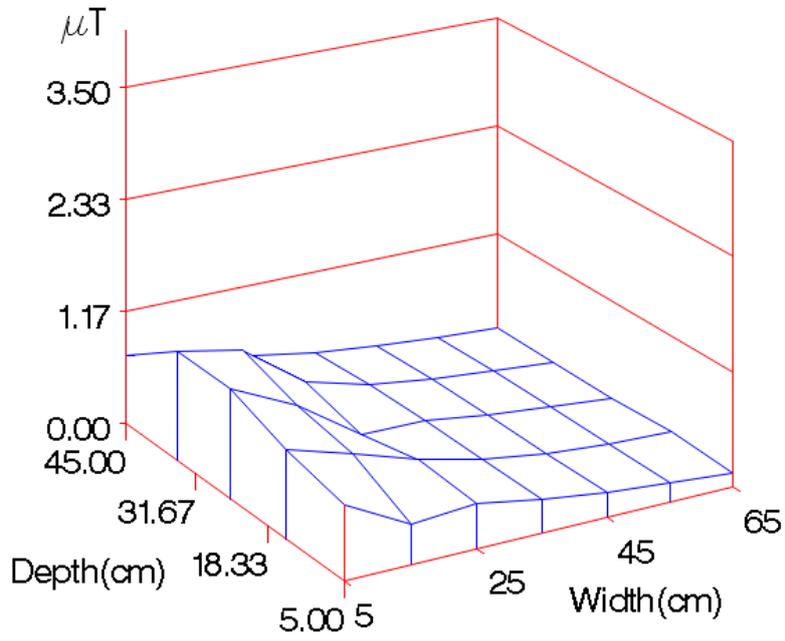

Integrated Magnetic Field At Height= 25cm

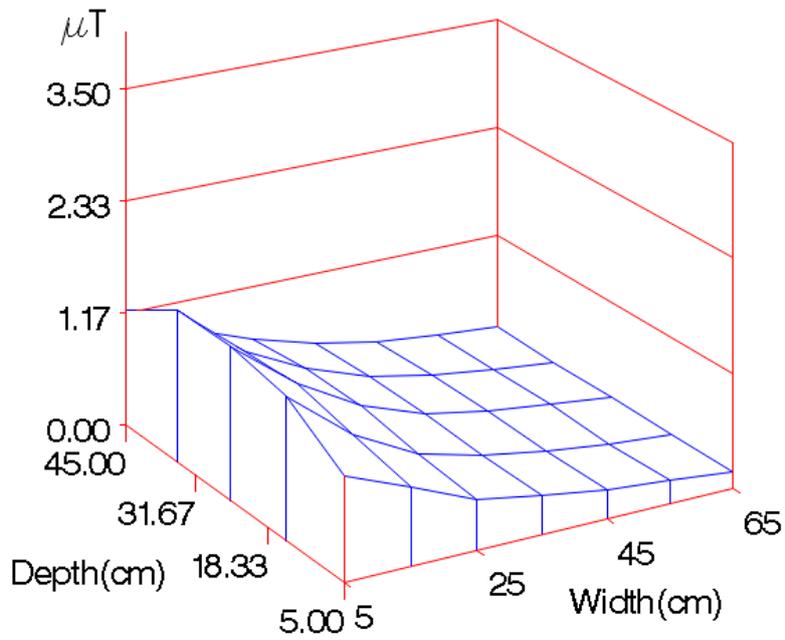

Integrated Magnetic Field At Height= 35cm

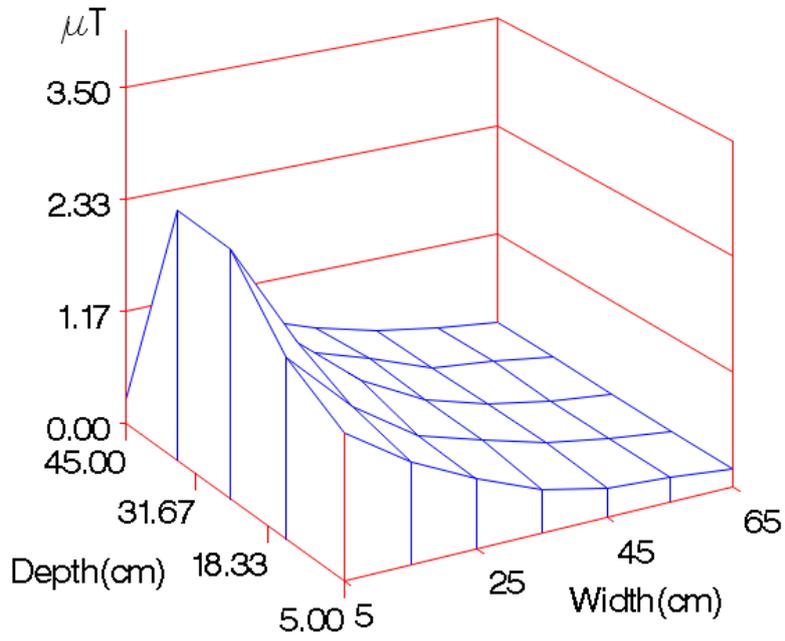

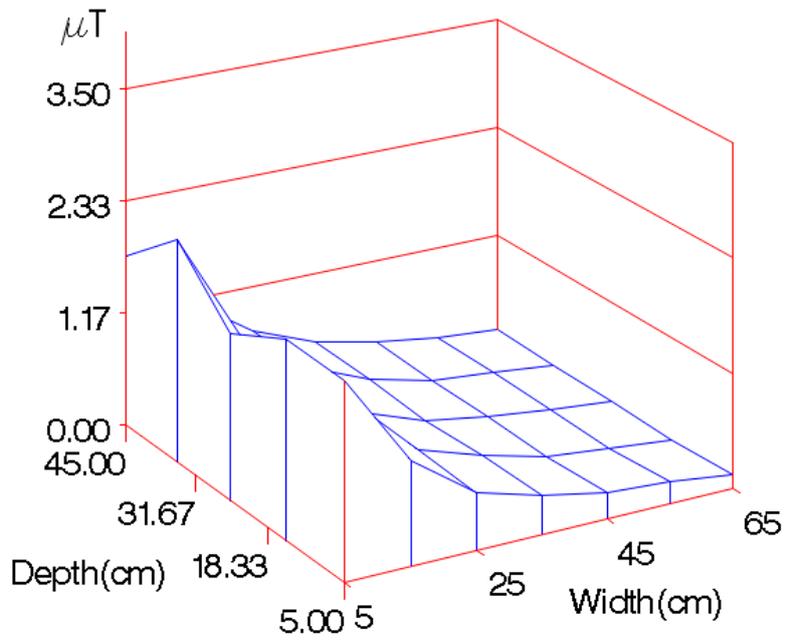

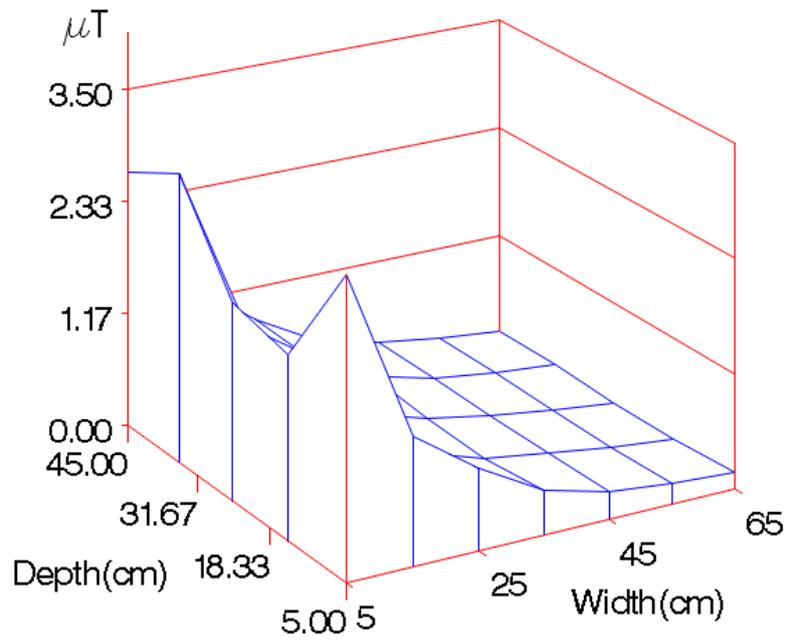